\documentclass{jfm}
\usepackage{float}
\usepackage{graphicx}
\usepackage{epstopdf, epsfig}
\usepackage{flushend}
\usepackage[dvipsnames]{xcolor}
\usepackage{amsmath}
\usepackage{latexsym}
\usepackage{caption}
\usepackage{color,soul}
\usepackage{physics}

\newcommand\St{\mbox{\textit{St}}}  % Strouhal number
\newcommand\ith{^{(i)}}
\newcommand\jth{^{(j)}}
\newcommand\ithH{^{(i)^*}}

\newcommand\qvec{\boldsymbol{q}}

\newcommand\qfft{\hat{\boldsymbol{q}}_{m\omega}}
\newcommand\ffft{\hat{\boldsymbol{f}}_{m\omega}}
\newcommand\yfft{\hat{\boldsymbol{y}}_{m\omega}}
\newcommand\qffthat{\tilde{\boldsymbol{q}}_{m\omega}}
\newcommand\fffthat{\tilde{\boldsymbol{f}}_{m\omega}}
\newcommand\yffthat{\tilde{\boldsymbol{y}}_{m\omega}}
\newcommand\resolvent{\mathsfbi{R}_{m\omega}}
\newcommand\simpleresolvent{\mathsfbi{H}_{m\omega}}
\newcommand\solidrule[1][0.5cm]{\rule[0.5ex]{#1}{1pt}}
\newcommand\dashedrule{\mbox{\solidrule[1mm]\hspace{1mm}\solidrule[1mm]\hspace{1mm}\solidrule[1mm]}}

\shorttitle{Guidelines for authors}
\shortauthor{O. Schmidt, A. Towne, G. Rigas, T. Colonius, G. Br\`{e}s}

\title{Spectral analysis of jet turbulence}

\author{Oliver T. Schmidt\aff{1}
\corresp{\email{oschmidt@caltech.edu}},
{Aaron Towne}\aff{2},
{Georgios Rigas}\aff{1},
{Tim Colonius}\aff{1},
\and Guillaume A. Br\`{e}s\aff{3}}

\affiliation{
\aff{1}California Institute of Technology, Pasadena, CA 91125, USA
\aff{2}Center for Turbulence Research, Stanford University, Stanford, CA 94305, USA
\aff{3}Cascade Technologies Inc., Palo Alto, CA 94303, USA
}

\begin{document}

\maketitle

\begin{abstract}
Informed by LES data and resolvent analysis of the mean flow, we examine the structure of turbulence in jets in the subsonic, transonic, and supersonic regimes. Spectral (frequency-space) proper orthogonal decomposition is used to extract energy spectra and decompose the flow into energy-ranked coherent structures. The educed structures are generally well predicted by the resolvent analysis. Over a range of low frequencies and the first few azimuthal mode numbers, these jets exhibit a low-rank response characterized by Kelvin-Helmholtz (KH) type wavepackets associated with the annular shear layer up to the end of the potential core and that are excited by forcing in the very-near-nozzle shear layer. These modes too the have been experimentally observed before and predicted by quasi-parallel stability theory and other approximations--they comprise a considerable portion of the total turbulent energy.  At still lower frequencies, particularly for the axisymmetric mode, and again at high frequencies for all azimuthal wavenumbers, the response is not low rank, but consists of a family of similarly amplified modes.  These modes, which are primarily active downstream of the potential core, are associated with the Orr mechanism.  They occur also as sub-dominant modes in the range of frequencies dominated by the KH response.  Our global analysis helps tie together previous observations based on local spatial stability theory, and explains why quasi-parallel predictions were successful at some frequencies and azimuthal wavenumbers, but failed at others. 
\end{abstract}


%%%%%%%%%%%%%%%%%%%%%%%%%%%%%%%%%%%%%%%%%%%%%%%%%%%%%%%%%%%%%%%%%%%%%%
\section{Introduction}
Large-scale coherent structures in the form of wavepackets play an important role in the dynamics and acoustics of turbulent jets. In particular, their spatial coherence makes wavepackets efficient sources of sound \citep{crighton1990shear,jordan2013wave}. They are most easily observed in forced experiments, where periodic exertion establishes a phase-reference. The measurements of a periodically forced turbulent jet by \citet{crow1971orderly} served as a reference case for wavepacket models. Early examples of such models include the studies by \citet{michalke1971instability} and \citet{crighton1976stability}, who established the idea that the coherent structures can be interpreted as linear instability waves evolving around the turbulent mean flow.

Wavepackets in unforced jets exhibit intermittent behavior \citep{cavalieri2011jittering} and as a result are best understood as statistical objects that emerge from the stochastic turbulent flow. We use spectral proper orthogonal decomposition \citep{Lumley:1970,TowneEtAt2017SPOD} to extract these structures from the turbulent flow. SPOD has been applied to a range of jets using data from both experimental \citep{glauser1987coherent,arndt1997proper,citriniti2000reconstruction,suzuki2006instability,gudmundsson2011instability}, and numerical studies \citep{sinha2014wavepacket,towne2015stochastic,SchmidtEtAl_2017_JFM}. 

Wavepackets have been extensively studied, mainly because of their prominent role in the production of jet noise, and models based on the parabolized stability equations (PSE) have proven to be successful at modeling them \citep{gudmundsson2011instability,cavalieri2013wavepackets,sinha2014wavepacket}. This agreement between SPOD modes and PSE solution breaks down at low frequencies and for some azimuthal wavenumbers, and in general downstream of the potential core. In the present study, we show that the SPOD eigenspectra unveil low-rank dynamics, and we inspect the corresponding modes and compare them with predictions based on resolvent analysis. The results show that two different mechanisms are active in turbulent jets, and they explain the success and failure of linear and PSE models.

Resolvent analysis of the turbulent mean flow field seeks sets of forcing and response modes that are optimal with respect to the energetic gain between them. When applied to the mean of a fully turbulent flow, the resolvent forcing modes can be associated with nonlinear modal interactions \citep{McKeonSharma2010} as well as stochastic inputs to the flow, for example the turbulent boundary layer in the nozzle that feeds the jet. \citet{garnaud2013pref} interpreted the results of their resolvent analysis of a turbulent jet in the light of experimental studies of forced jets by \citet{moore1977role} and \citet{crow1971orderly}. They found that the frequency of the largest gain approximately corresponds to what the experimentalists referred to as the \emph{preferred frequency}, i.e.~the frequency where external harmonic forcing in the experiments triggered the largest response. In the context of jet aeroacoustics, \citet{jeun2016input} restricted the optimal forcing to vortical perturbations close to the jet axis, and the optimal responses to the far-field pressure. Their results show that suboptimal modes have to be considered in resolvent-based jet noise models.

In this paper, we use resolvent analysis to model and explain the low-rank behavior of turbulent jets revealed by SPOD.  Recent theoretical connections between SPOD and resolvent analysis \citep{towne2015stochastic,lesshafftmodeling,TowneEtAt2017SPOD} make the latter a natural tool for this endeavor and provide a framework for interpreting our results.

The remainder of the paper is organized as follows. The three large-eddy simulation databases used to study different Mach number regimes are introduced in \S \ref{LES}. At first, the focus is on the lowest Mach number case in \S\S \ref{SPOD}-\ref{comparison}. We start by analyzing the data of this case using (mainly) SPOD in \S \ref{SPOD}, followed by the resolvent analysis in \S \ref{resolvent}. The results of the SPOD and the resolvent model are compared in \S \ref{comparison}. \S \ref{mach} addresses Mach number effects observed in the remaining two cases representing the transonic and the supersonic regime. Finally, the results are discussed in \S \ref{discussion}. For completeness, we report in appendix \ref{app:mach} all results for the higher Mach number cases that were omitted in \S \ref{mach} for brevity.

%%%%%%%%%%%%%%%%%%%%%%%%%%%%%%%%%%%%%%%%%%%%%%%%%%%%%%%%%%%%%%%%%%%%%%
\section{Large eddy simulation}\label{LES}
%%%%%%%%%%%%%%%%%%%%%%%%%%%%%%%%%%%%%%%%%%%%%%%%%%%%%%%%%%%%%%%%%%%%%%

The unstructured flow solver ``Charles" \citep{bres2017unstructured} is used to perform large-eddy simulations of three turbulent jets at Mach numbers $M_j=U_j/a_j$ of 0.4, 0.9, and 1.5. All jets are isothermal and the supersonic jet is ideally expanded. The nozzle geometry is included in the computational domain, and synthetic turbulence combined with a wall model is applied inside the nozzle to obtain a fully turbulent boundary layer inside the nozzle. The jets are further characterized by their Reynolds numbers $\Rey=\rho_jU_jD/\mu_j$, which correspond approximately to the laboratory values in a set of companion experiments. The reader is referred to \citet{bres2017unstructured} for further details on the numerical method, meshing strategy and subgrid-scale model. A detailed validation of the $M_j=0.9$ jet including the nozzle-interior turbulence modeling (i.e., synthetic turbulence, wall model) can be found in \citet{bres2017jets}. The subscripts $j$ and $\infty$ refer to jet and free-stream  conditions, $a$ is the speed of sound, $\rho$ density, $D$ nozzle diameter, $\mu$ dynamic viscosity, $T$ temperature, and $U_j$ to the axial jet velocity on the centerline of the nozzle exit, respectively. Throughout this paper, the flow is non-dimensionalized by its nozzle exit values, pressure by $\rho_j U_j^2$, lengths by $D$, and time by $D/U_j$. Frequencies are reported in terms of the Strouhal number $\St=\omega/(2\pi M_j)$, where $\omega$ is the non-dimensional angular frequency.
\begin{table}
  \begin{center}
\def~{\hphantom{0}}
  \begin{tabular}{lccccccc}
  %\multicolumn{8}{ c }{Large eddy simulation} \\ 
      case  & $M_j$ & $\Rey$ & $\frac{p_0}{p_\infty}$ & $\frac{T_0}{T_\infty}$ & $n_{\text{cells}}$ & $\frac{\dd t c_\infty}{D}$ & $\frac{t_{\text{sim}} c_\infty}{D}$ \\[3pt]
      {subsonic} & 0.4 &  $0.45\cdot10^6$ & 1.117 & 1.03 & $15.9\cdot10^6$  & $1\cdot10^{-3}$ & 2000\\
      {transsonic} & 0.9 & $1.01\cdot10^6$ & 1.7 & 1.15 & $15.9\cdot10^6$  & $1\cdot10^{-3}$ & 2000\\
      {supersonic} & 1.5 & $1.76\cdot10^6$ & 3.67 & 1.45 & $31\cdot10^6$  & $2.5\cdot10^{-4}$ & 500
  \end{tabular}
  \caption{Parameters of the large-eddy simulations.}
  \label{tab:LES}
  \end{center}
\end{table}
The parameters of the three simulations are listed in table \ref{tab:LES}: ${p_0}/{p_\infty}$ is the nozzle pressure ratio, ${T_0}/{T_\infty}$ the nozzle temperature ratio, $n_{\text{cells}}$ the number of control volumes, ${\dd t c_\infty}/{D}$ the computational time step, and ${t_{\text{sim}}/c_\infty}{D}$ the total simulation time after the flow became stationary, i.e.~without initial transients. 
\begin{figure}
\centering
	\includegraphics[trim=0 2.3cm 0 3mm, clip, width=1\textwidth]{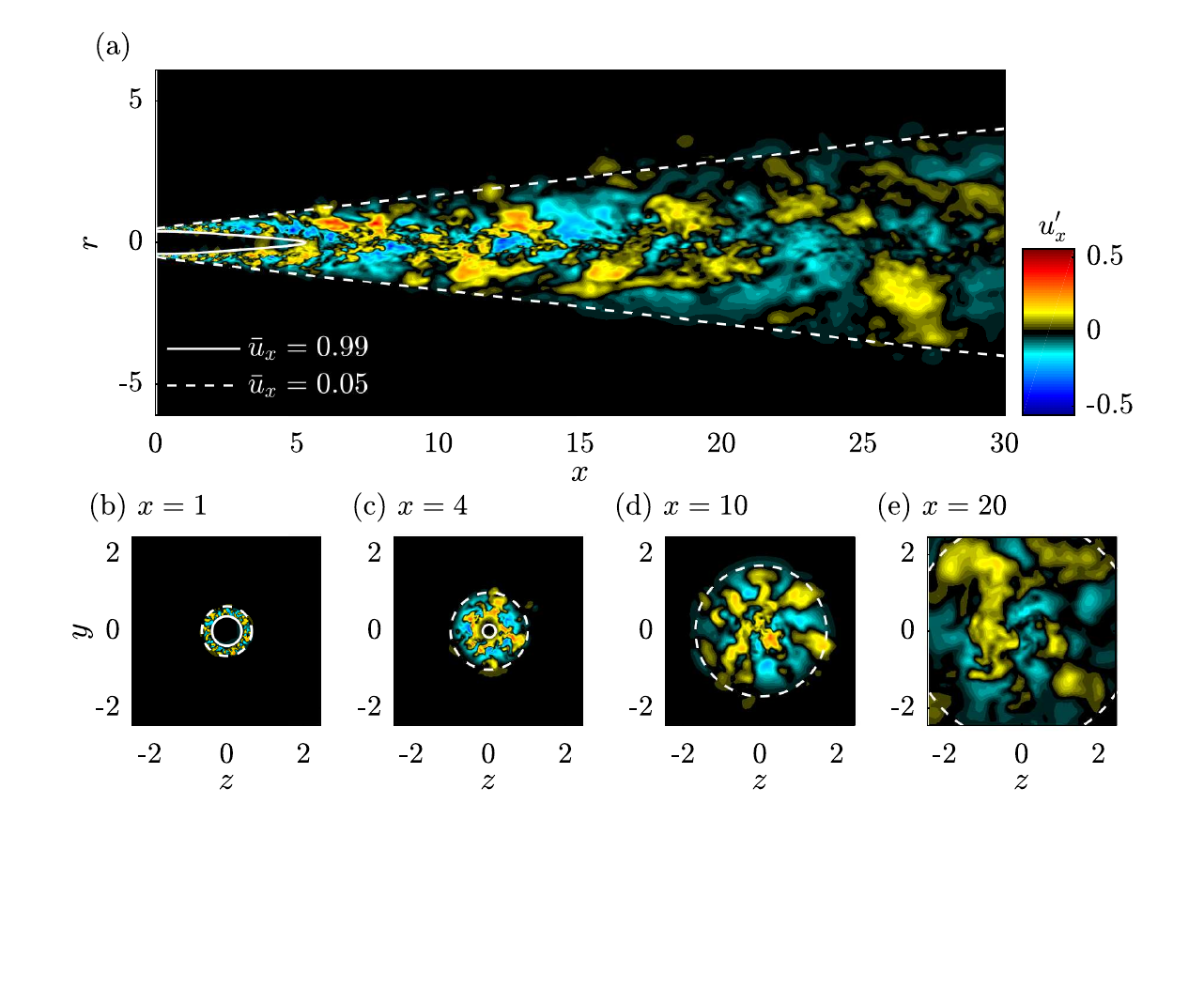}
	\captionsetup{singlelinecheck=off}
	\caption[]{Instantaneous flow field of the subsonic jet: (a) streamwise cross-section along the jet axis; (b-e) transverse planes at different streamwise locations $x$. The streamwise velocity fluctuation ${u}'_x$ is shown. The potential core and the jet width are indicated as lines of constant $\bar{u}_x$ at 99\% and 5\% of the jet velocity $U_j$, respectively.}
	\label{fig:M04_LESinst}
\end{figure}
The unstructured LES data is first interpolated onto a $n_x\times n_r \times n_\theta$ structured cylindrical grid spanning $x,r,\theta\in[0,\,30]\times[0,\,6]\times[0,\,2\pi]$. Snapshots are saved with a temporal separation ${\Delta t c_\infty}/{D}$ (in acoustical units). For both the spectral analysis in \S \ref{SPOD}, and the resolvent model in \S \ref{resolvent}, we Reynolds decompose a flow quantity $q$ into the long-time mean denoted by $\overline{(\cdot)}$ and the fluctuating part ${(\cdot)'}$ as
\begin{equation}\label{eqn:reynbolds}
q(x,r,\theta,t)=\overline{q}(x,r,\theta) + q'(x,r,\theta,t).
\end{equation}
A visualization of the subsonic jet is shown in figure \ref{fig:M04_LESinst}. Only the domain of interest for this study is shown  (the full LES domain is much larger and includes flow within the nozzle as well as far-field sponge regions).

\begin{figure}
\centering
	\includegraphics[trim=0 0.6cm 0 3mm, clip, width=1\textwidth]{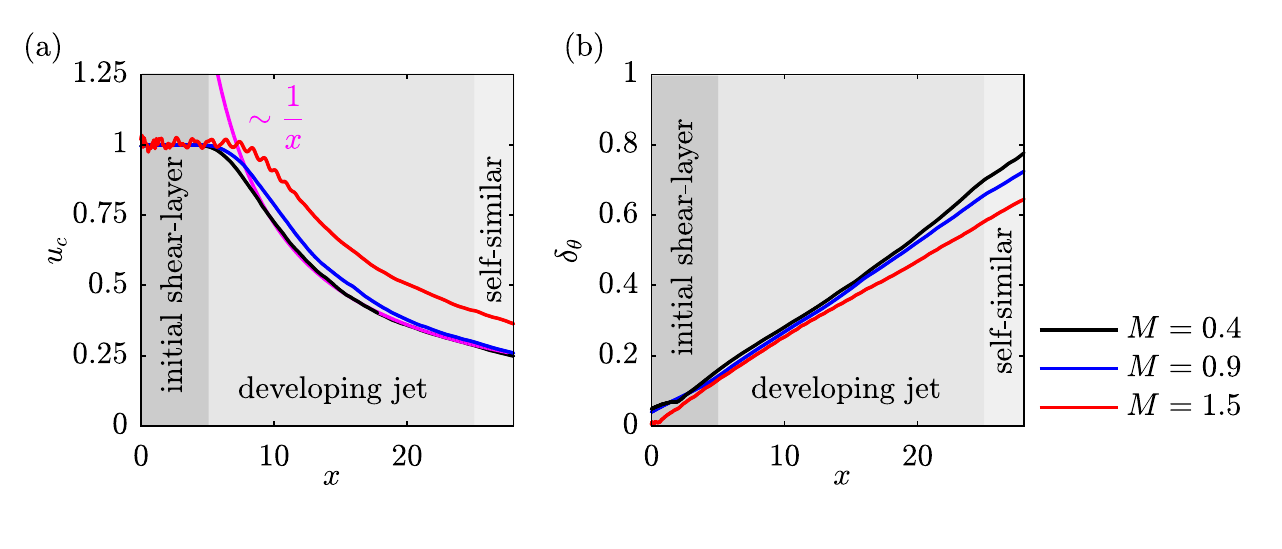}
	\captionsetup{singlelinecheck=off}
	\caption[]{Mean flow characteristics for the three jets: (a) centerline velocity $u_{c}=\overline{u}_x(x,r=0)$; (b) momentum thickness $\delta_{\theta}=\int_0^{r_{1}} \frac{\bar{\rho}\bar{u}_x}{\rho_c u_c} \left( 1-\frac{\bar{u}_x}{u_c} \right) \dd r$, where the integral is taken from the centerline to $r_{1}$ defined as $\bar{u}_x(r_{1})=\bar{u}_\infty+0.01\bar{u}_{j}$  (the $\bar{u}_{\infty}$ term is included to account for a small coflow included in the simulations). The centerline velocity becomes inversely proportional to the axial distance shortly after the potential core, and the momentum thickness increases approximately linearly.}
	\label{fig:baseFlow_ucl_r12}
\end{figure}
The mean centerline velocity of the three jets is compared in figure \ref{fig:baseFlow_ucl_r12}(a). The plateau close to the nozzle characterizes the potential core, whose length increases with Mach number. A weak residual shock pattern is observed for $x\lesssim10$ in the supersonic case. The streamwise development of turbulent jets is usually described in terms of an \emph{initial development region} ($0\lesssim x/D \lesssim 25$), and a \emph{self-similar region} ($x/D \gtrsim 25$), see e.g.~\cite{Pope2000}. In the latter, the jet is fully described by a self-similar velocity profile, and a constant spreading and velocity-decay rate. For a dynamical description of the flow, we further divide the initial development region into two parts. The \emph{initial shear-layer region} extends up to the end of the potential core ($0\lesssim x/D \lesssim 5$ for $M=0.4$) and is characterized by a constant velocity jump over the shear-layer, and a linearly increasing shear-layer thickness. The \emph{developing jet region} lies between the end of the potential core and the start of the self-similar region ($5\lesssim x/D \lesssim 25$). In this region, the centerline velocity transitions rapidly to its asymptotic decaying behavior. The mean radial velocity profile, on the other hand, has not yet converged to its self-similar downstream solution. The $1/x$-scaling of the centerline velocity with streamwise distance \citep{Pope2000} is indicated for the subsonic case. The momentum thickness shown in figure \ref{fig:baseFlow_ucl_r12}(b) increases approximately linear in all three regions in all cases. These characteristic velocities and length scales in each region imply different frequency scalings that will become important later.

\begin{table}
  \begin{center}
\def~{\hphantom{0}}
  \begin{tabular}{lccccccc|ccc}
  \multicolumn{8}{ c }{Interpolated database}  & \multicolumn{3}{ c }{SPOD}\\ 
      case & $\frac{\Delta t c_\infty}{D}$ & $n_x$ & $n_r$ & $n_\theta$ & $x_1$ & $r_1$ & $n_{t}$ & $n_{\text{freq}}$& $n_{\text{ovlp}}$& $n_{\text{blk}}$\\[3pt]
      {subsonic} & 0.2 & 656 & 138 & 128 & 30 & 6 & 10,000 & 256 & 128 & 78 \\
      {transsonic} & 0.2 & 656 & 138 & 128 & 30 & 6 & 10,000 & 256 & 128 & 78  \\
      {supersonic} & 0.1 & 698 & 136 & 128 & 30 & 6 & 5,000 & 256 & 128 & 39 
  \end{tabular}
  \caption{Parameters of the structured cylindrical grid of the interpolated database (left), and the SPOD parameters (right).}
  \label{tab:SPOD}
  \end{center}
\end{table}

In \S\S \ref{SPOD}-\ref{comparison}, we will focus on the $M=0.4$ jet as the main conclusions are similar for all three Mach number regimes, and we address Mach number effects in detail in \S \ref{mach}.

%%%%%%%%%%%%%%%%%%%%%%%%%%%%%%%%%%%%%%%%%%%%%%%%%%%%%%%%%%%%%%%%%%%%%%%%%%%%%%%%%%%%  
\section{Spectral analysis of the LES data}\label{SPOD}
%%%%%%%%%%%%%%%%%%%%%%%%%%%%%%%%%%%%%%%%%%%%%%%%%%%%%%%%%%%%%%%%%%%%%%%%%%%%%%%%%%
In this section, we use spectral proper orthogonal decomposition (SPOD) to identify coherent structures within the three turbulent jets. This form of proper orthogonal decomposition (POD) identifies energy-ranked modes that each oscillate at a single frequency, are orthogonal to other modes at the same frequency, and, as a set, optimally represent the space-time flow statistics. SPOD was introduced by \citet{Lumley:1967,Lumley:1970} but has been used sparingly compared to the common spatial form of POD \citep{sirovich1987turbulence,aubry1991hidden} and dynamic mode decomposition \citep{schmid2010dmd}.  However, recent work by \cite{TowneEtAt2017SPOD} showed that SPOD combines the advantages of these other two methods--SPOD modes represent coherent structures that are dynamically significant and optimally account for the random nature of turbulent flows.  This makes SPOD an ideal tool for identifying coherent structures within the turbulent jets considered in this paper.

We seek modes that are orthogonal in the inner product
\begin{equation}\label{eqn:norm}
\left<\qvec_1,\qvec_2\right>_E= \iiint \qvec_1^* \mathrm{diag} \left( 
\frac{\overline{T}}{\gamma\overline{\rho}M^2},
\overline{\rho},
\overline{\rho},
\overline{\rho},
\frac{\overline{\rho}}{\gamma(\gamma-1)\overline{T}M^2}
\right) \qvec_2 r \dd x \dd r \dd\theta,
\end{equation}
which are optimal in an induced compressible energy norm $\left<\cdot,\cdot\right>_E$ \citep{chu1965energy}. The energy weights are defined for the state vector $\qvec=[\rho, u_x, u_r, u_\theta, T]^T(x,r,\theta,t)$ of primitive variables density $\rho$, cylindrical velocity components $u_x$, $u_r$ and $u_\theta$, and temperature $T$. The notation $(\cdot)^*$ indicates the Hermitian transpose. We discretize the inner product defined by equation (\ref{eqn:norm}) as
\begin{equation}\label{eqn:discnorm}
\left<\qvec_1,\qvec_2\right>_E=\qvec_1^*\mathsfbi{W}\qvec_2,
\end{equation}
where the weight matrix $\mathsfbi{W}$ accounts for both numerical quadrature weights and the energy weights. Since the jet is stationary and symmetric with respect to rotation about the jet axis, it can be decomposed into azimuthal Fourier modes $\hat{(\cdot)}_m$ of azimuthal wavenumber $m$, temporal Fourier modes $\hat{(\cdot)}_\omega$ of angular frequency $\omega$, or combined spatio-temporal Fourier modes  Fourier modes $\hat{(\cdot)}_{m\omega}$ as
\begin{equation}\label{eqn:fft}
\qvec(x,r,\theta,t) = \sum_{m} \hat{\boldsymbol{q}}_{m}(x,r,t)\mathrm{e}^{\mathrm{i}m\theta} = \sum_{\omega} \hat{\boldsymbol{q}}_{\omega}(x,r,\theta)\mathrm{e}^{\mathrm{i}\omega t} = \sum_{m}\sum_{\omega} \qffthat(x,r)\mathrm{e}^{\mathrm{i}m\theta}\mathrm{e}^{\mathrm{i}\omega t}.
\end{equation}
To calculate the SPOD, the data is first segmented into sequences $\boldsymbol{Q}=\left[\qvec^{(1)}\;\qvec^{(2)}\cdots\qvec^{(n_\text{freq})} \right]$ each containing $n_{\text{freq}}$ instantaneous snapshots of $\qvec$ which are considered to be statistically independent realizations of the flow under the ergodic hypothesis. Details on the interpolated databases and spectral estimation parameters are listed in table \ref{tab:SPOD}. The spatio-temporal Fourier transform of the $l$-th block yields $\hat{\boldsymbol{Q}}_{m\omega_k}^{(l)}=\left[\hat{\boldsymbol{q}}_{m\omega_{1}}^{(l)}\;\hat{\boldsymbol{q}}_{m\omega_{2}}^{(l)}\cdots\hat{\boldsymbol{q}}_{m\omega_{n_{\text{freq}}}}^{(l)} \right]$, where $\hat{\boldsymbol{q}}_{m\omega_{k}}^{(l)}$ is the $l$-th realization of the Fourier transform at the $k$-th discrete frequency. A periodic Hann window is used to minimize spectral leakage. The ensemble of $n_{\text{blk}}$ Fourier realizations of the flow at a given frequency $\omega_k$ and azimuthal wavenumber $m$ are now collected into a data matrix $\hat{\boldsymbol{Q}}_{m\omega_{k}}=\left[\hat{\boldsymbol{q}}_{m\omega_{k}}^{(1)}\;\hat{\boldsymbol{q}}_{m\omega_{k}}^{(2)}\cdots\hat{\boldsymbol{q}}_{m\omega_{k}}^{(n_{\text{blk}})} \right]$. For a particular $m$ and $\omega_k$, the SPOD modes are found as the eigenvectors $\boldsymbol{\Psi}_{m\omega_{k}}=\left[{\boldsymbol{\psi}}_{m\omega_{k}}^{(1)}\;{\boldsymbol{\psi}}_{m\omega_{k}}^{(2)}\cdots{\boldsymbol{\psi}}_{m\omega_{l}}^{(n_{\text{blk}})} \right]$, and the modal energy as the corresponding eigenvalues $\boldsymbol{\Lambda}_{m\omega_k}=\text{diag}\left(\lambda_{m\omega_k}^{(1)},\lambda_{m\omega_k}^{(2)},\cdots,\lambda_{m\omega_k}^{(n_{\text{blk}})}\right)$ of the weighted cross-spectral density matrix $\hat{\boldsymbol{S}}_{m\omega_{k}}=\hat{\boldsymbol{Q}}_{m\omega_{k}}\hat{\boldsymbol{Q}}_{m\omega_{k}}^*$ as
\begin{equation}
\hat{\boldsymbol{S}}_{m\omega_{k}}\boldsymbol{W}\boldsymbol{\Psi}_{m\omega_{k}}=\boldsymbol{\Psi}_{m\omega_{k}} \boldsymbol{\Lambda}_{m\omega_{k}}.
\end{equation}
The modes are sorted by decreasing energy, i.e.~$\lambda_{m\omega_k}^{(1)}\geq\lambda_{m\omega_k}^{(2)}\geq\cdots\geq\lambda_{m\omega_k}^{(n_{\text{blk}})}$.
This formulation guarantees that the SPOD modes have the desired orthonormality property $\left<\boldsymbol{\psi}_{m\omega_{k}}^{(i)},\boldsymbol{\psi}_{m\omega_{k}}^{(j)}\right>_E=\delta_{ij}$, where $\delta_{ij}$ is the Kronecker delta function, and are optimal in terms of modal energy in the norm induced by equation (\ref{eqn:discnorm}). For brevity, we denote the $l$-th eigenvalue and the pressure component of the corresponding eigenmode as $\lambda_{l}$ and $\psi_{p}^{(l)}$, respectively. The dependence on a specific azimuthal wavenumber and frequency is implied and given in the description. Since the SPOD modes are optimal in terms of energy, we sometimes refer to the first SPOD mode as the \emph{leading} or \emph{optimal} mode and to the subsequent lower-energy modes as \emph{suboptimal} modes.

Since we wish to express the data in terms of modes that oscillate at real and positive frequencies, we take the temporal Fourier transform in equation (\ref{eqn:fft}) first. Statistical homogeneity in $\theta$ implies that averaged quantities are the same for any $\pm m$.  After verifying statistical convergence, we add the contributions of positive and negative $m$. 

The distribution of power into the frequency components a signal is comprised of is referred to as its power spectrum. Power spectra are commonly expressed in terms of the power spectral density (PSD), which we will introduce later in equation (\ref{eqn:psd}). In the context of SPOD, we are interested in finding a graphical representation that can be interpreted in a similar way. In each frequency bin, the discrete SPOD spectrum is represented by the decreasing energy levels of the corresponding set of eigenfunctions. There is no obvious continuity in the modal structure of the most energetic mode (or any other) from one frequency bin to the next. Nevertheless, it is instructive to examine how the modal energy changes as a function of frequency, so that in what follows we plot the SPOD eigenvalues for each mode $l$ as functions $\lambda_{l}(\St)$ of frequency and refer to the resulting set of curves as the \emph{SPOD eigenvalue spectrum}. 

\begin{figure}
	\centerline{\includegraphics[trim=0 1mm 0 6mm, clip, width=1\textwidth]{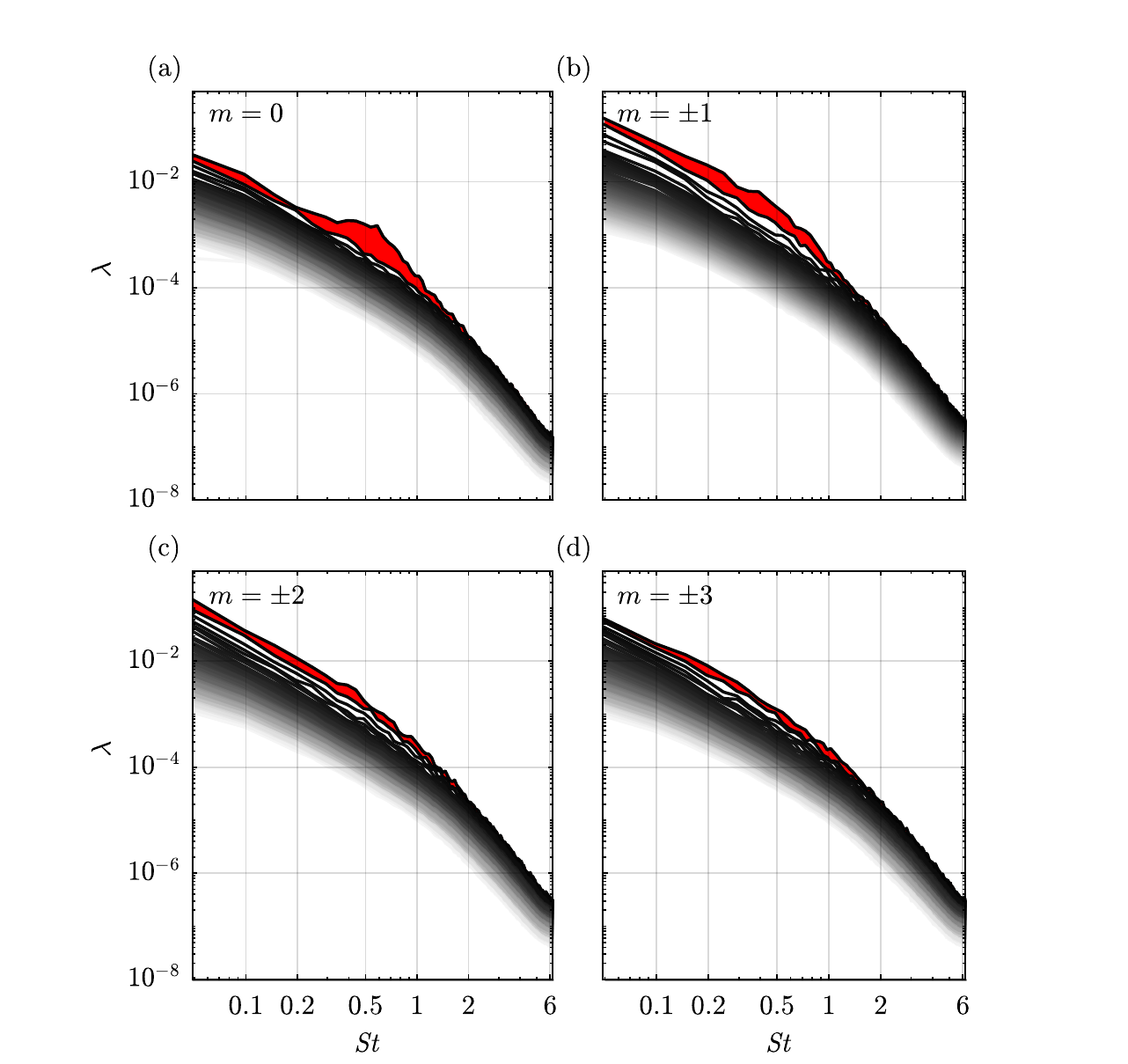}}% Images in 100% size
	\captionsetup{singlelinecheck=off}
	\caption[]{SPOD eigenvalue spectra ($\blacksquare\!{\color{Gray}\blacksquare}\!\square$, $\lambda_1\!>\!\lambda_2\!>\!\cdots\!>\!\lambda_{N}$) for the subsonic jet: (a) $m=0$; (b) $m=1$; (c) $m=2$; (d) $m=3$. For, $m>0$, the positive and negative azimuthal wavenumber components are summed. Red shaded areas (${\color{Red}\blacksquare}$) highlight the separation between the first and second modes.}
	\label{fig:SPODvsSPODforcing_spectrum_m01}
\end{figure}
The SPOD eigenvalue spectrum of the axisymmetric component of the subsonic jet is shown in figure \ref{fig:SPODvsSPODforcing_spectrum_m01}(a). The red shaded area highlights the separation between the first and second modes. A large separation indicates that the leading mode is significantly more energetic than the others. When this occurs, the physical mechanism associated with the first mode is prevalent, and we say that the flow exhibits \emph{low-rank behavior}.

The low-rank behavior is apparent over the frequency band $0.2\lesssim\St\lesssim2$, and peaks at $\St\approx0.6$. It is most pronounced for $m=0$ and $m=1$, shown in panel \ref{fig:SPODvsSPODforcing_spectrum_m01}(a) and \ref{fig:SPODvsSPODforcing_spectrum_m01}(b), respectively. With increasing azimuthal wavenumber, the low-rank behaviour becomes less and less pronounced. For $m=1$, the dominance of the optimal gain persists to low frequencies, whereas it cuts off below $\St\approx0.2$ for $m=0$. For $m=2$ in figure \ref{fig:SPODvsSPODforcing_spectrum_m01}(c), even though the eigenvalue separation is not very large, it is clear that the first resolvent mode at $St=0.2$ is a continuation of the same mode at higher frequencies. In contrast, the leading $m=0$ mode at $St=0.2$ appears to be a continuation of underlying band of the subpotimal SPOD modes. Since a global integral energy norm is used, it is important to note that the structure of the modes and their truncation by the domain unavoidably factor into the form of the energy spectra. These results should be compared with the optimal gain spectra shown in figure \ref{fig:M04_gainVsSt_m0123}. The corresponding spectra for the transsonic and supersonic cases are reported in appendix \ref{app:mach} (figures \ref{fig:M09_POD_spectrum_m0123} and \ref{fig:B118_POD_spectrum_m0123}).

\begin{figure}
	\centerline{\includegraphics[trim=0 6.2cm 0 3mm, clip, width=1\textwidth]{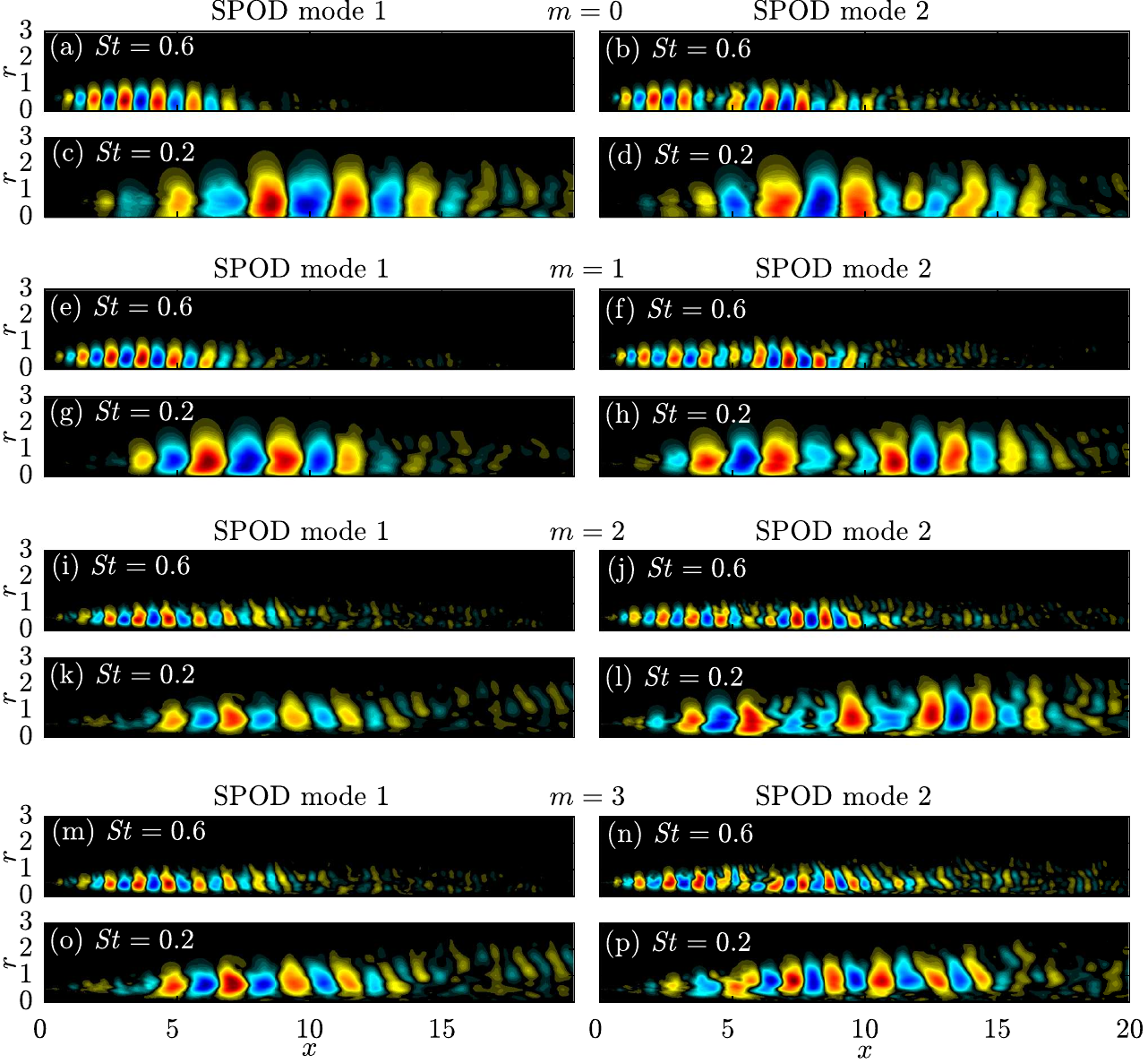}}% Images in 100% size
	\captionsetup{singlelinecheck=off}
	\caption[]{Comparison between SPOD mode 1 (left) and SPOD mode 2 (right) at two representative frequencies for the subsonic jet: (a-d) $m=0$; (e-h) $m=1$; (i-l) $m=2$; (m-p) $m=3$. The normalized pressure component (${\color{Red}\blacksquare}\!{\color{Black}\blacksquare}\!{\color{Blue}\blacksquare}$, $-1<\psi_p/\|\psi_p\|_\infty<1$) is shown in $x,r\in[0\;20]\times[0\;3]$.}
	\label{fig:M04_SPODmodes_m013}
\end{figure}
Figure \ref{fig:M04_SPODmodes_m013} shows the first (left column) and the second (right column) SPOD modes  at two representative frequencies and for $m=0,\dots,3$. The most energetic mode at $m=0$ and $\St=0.6$ is shown in figure \ref{fig:M04_SPODmodes_m013}(a). These parameters correspond approximately to the point of maximum separation between the first and second mode in figure \ref{fig:SPODvsSPODforcing_spectrum_m01}(a). The pressure field takes the form of a compact wavepacket in the initial shear-layer region of the jet (see figure \ref{fig:baseFlow_ucl_r12}). Its structure is reminiscent of the Kelvin-Helmholtz shear-layer instability of the mean flow \citep{suzuki2006instability,gudmundsson2011instability,jordan2013wave}. \citet{crighton1976stability} argued that the mean flow can be regarded as an equivalent laminar flow, and found that it is convectively unstable (in the local weakly non-parallel sense) in the initial shear-layer region. Following their interpretation of this structure as a modal spatial instability wave, we refer to it as a \emph{KH-type wavepacket}. The suboptimal SPOD mode shown in \ref{fig:M04_SPODmodes_m013}(b) has a double-wavepacket structure with an upstream wavepacket similar to the KH-type and a second wavepacket further downstream. This double-wavepacket structure is similarly observed at other frequencies and azimuthal wavenumbers, most prominently in figure \ref{fig:M04_SPODmodes_m013}(b,f,h,j,n). The turbulent mean flow in this region is convectively stable--it does not support spatial modal growth. \citet{tissot2017} argue that the presence of large-scale structures in this region can be explained in terms of non-modal growth through the Orr mechanism. Our resolvent model presented in \S \ref{resolvent} supports this idea, and we therefore term these \emph{downstream} or \emph{Orr-type} wavepackets. Large parts of this paper are dedicated to establish a clear separation and explanation of these two distinct mechanisms.

\begin{figure}
	\centerline{\includegraphics[trim=0 3.6cm 0 5mm, clip, width=1\textwidth]{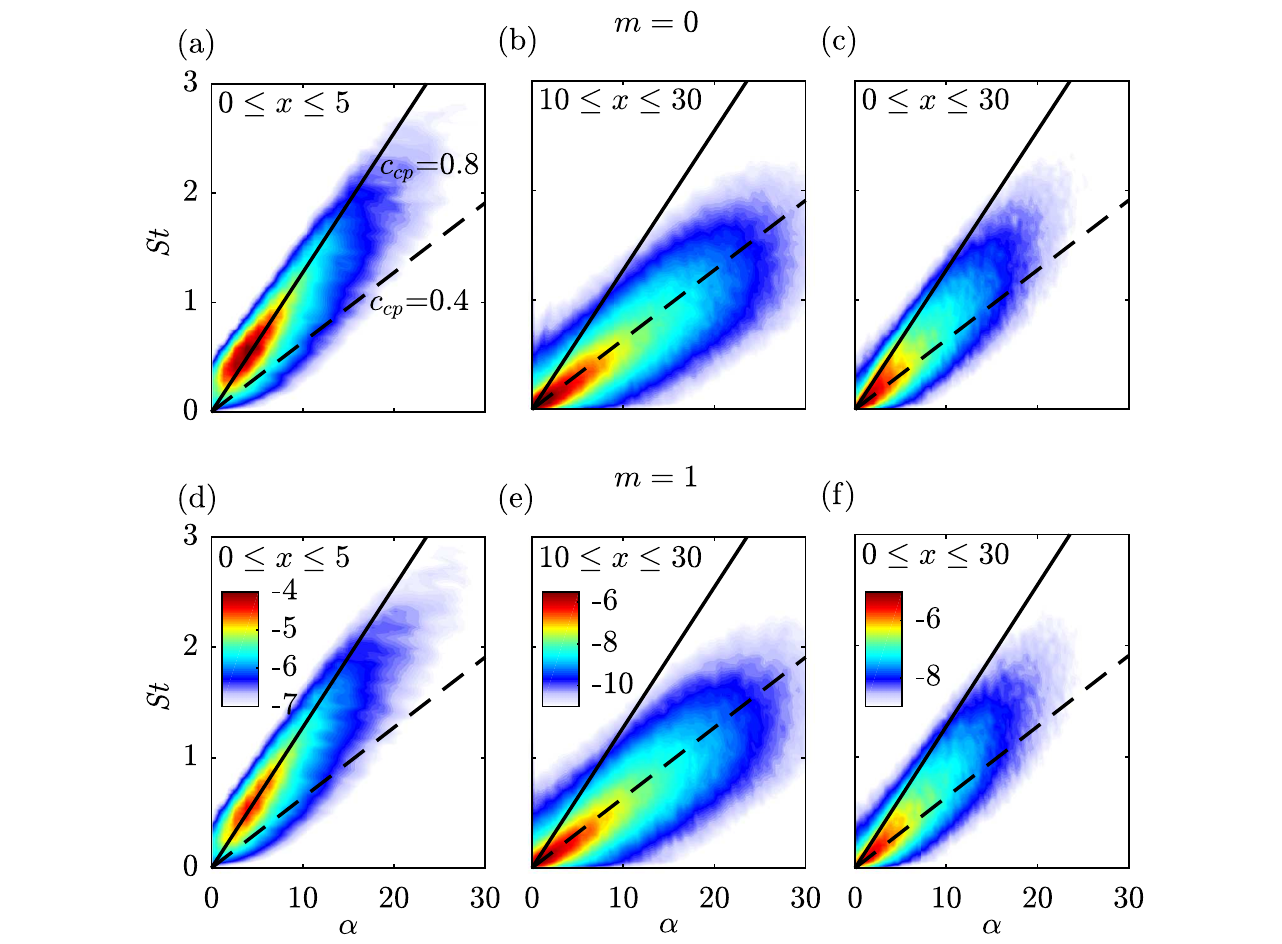}}% Images in 100% size
	\captionsetup{singlelinecheck=off}
	\caption[]{Frequency-wavenumber diagrams ($\square\!{\color{Cyan}\blacksquare}\!{\color{Red}\blacksquare}$, $\log_{10}(\bar{P}_{pp})$) of the subsonic simulation along the lip-line ($r=0.5$) over axial sections representative of the potential core (left column), downstream of the potential core (center column), and the entire jet (right column). The behaviors for $m=0$ (top row) and $m=1$ (bottom row) are similar. Two lines of constant phase speed ({\solidrule}, $c_{cp}=0.8$) and ({\dashedrule}, $c_{cp}=0.4$) are shown as a reference. Rows share the same contour levels.}
	\label{fig:M04_StvsAlpha_SPOD}
\end{figure}

In order to further isolate mechanisms associated with different regions of the jet, we construct empirical frequency-wavenumber diagrams by taking the Fourier transform in the streamwise direction of the LES pressure along the lip line ($r=0.5$). We compute the spatio-temporal PSD as
\begin{equation} \label{eqn:psd}
\bar{P}_{qq} = \frac{1}{n_{\text{blk}}}\sum_{l=1}^{n_{\text{blk}}} \left|\hat{q}^{(l)}_{m\omega\alpha}\right|^2, 
\end{equation}	
where $\hat{(\cdot)}_\alpha$ is the Fourier transform in the axial direction and $\alpha$ the axial wavenumber. In figure \ref{fig:M04_StvsAlpha_SPOD}, the PSD is plotted, and compared to the PSD computed with different window functions constraining the signal to specific regions along the streamwise axis, specifically $0 \le x \le 5$, representing the annular shear layer, and $10 \le x < 30$, the developing jet. 

Qualitatively  similar results are found for $m=0$ (top row) and $m=1$ (bottom row). In the initial shear layer region investigated in figure \ref{fig:M04_StvsAlpha_SPOD}a and \ref{fig:M04_StvsAlpha_SPOD}d, the pressure PSD follows the line of constant phase speed $c_{ph}=0.8$ and peaks at $\St\approx0.5$ for $m=0$, and a slightly lower frequency for $m=1$. This phase speed is typical for KH-type shear-layer instability waves, and the peak frequencies are close to the ones where the low-rank behavior is most pronounced in figure \ref{fig:SPODvsSPODforcing_spectrum_m01}a and \ref{fig:SPODvsSPODforcing_spectrum_m01}b, respectively. In the developing jet region in figure \ref{fig:M04_StvsAlpha_SPOD}b and \ref{fig:M04_StvsAlpha_SPOD}e, waves propagate with about half of the phase speed observed in the initial shear-layer, and the PSD peaks at the lowest resolved frequency. The PSD for the entire domain shown in figure \ref{fig:M04_StvsAlpha_SPOD}c and \ref{fig:M04_StvsAlpha_SPOD}f, by construction, combines these effects.

%%%%%%%%%%%%%%%%%%%%%%%%%%%%%%%%%%%%%%%%%%%%%%%%%%%%%%%%%%%%%%%%%%%%%%%%%%%%%%%%%%%%
\section{Resolvent model}\label{resolvent}
%%%%%%%%%%%%%%%%%%%%%%%%%%%%%%%%%%%%%%%%%%%%%%%%%%%%%%%%%%%%%%%%%%%%%%%%%%%%%%%%%%%%
A key concept that emerged from the early studies of transient growth \citep{farrell1993stochastic,trefethen1993hydrodynamic,reddy1993pseudospectra,reddy1993energy} is that of the resolvent operator. The resolvent operator is derived from the forced linearized equations of motion and constitutes a transfer function between body forces and corresponding responses. It has been used to study the linear response to external forcing of a range of laminar flows including channel flow \citep{jovanovic2005componentwise,moarref2012model}, boundary layers \citep{monokrousos2010global,sipp2012characterization} and the flow over a backward-facing step \citep{dergham2013stochastic}. 

When the resolvent is computed for the turbulent mean flow, the forcing can be identified with the nonlinear effects, namely the triadic interactions that conspire to force a response at a given frequency and azimuthal wavenumber \citep{McKeonSharma2010}. Previous resolvent analyses of turbulent jets include those of jets \citet{garnaud2013pref}, \citet{jeun2016input} and \citet{lesshafftmodeling}.

The use of a resolvent model is motivated by recent findings that connect SPOD and resolvent analysis \citep{towne2015stochastic,lesshafftmodeling,TowneEtAt2017SPOD}. Specifically, SPOD and resolvent modes are identical when the SPOD expansion coefficients are uncorrelated, which is typically associated with white-noise forcing.

%%%%%%%%%%%%%%%%%%%%%%%%%%%%%%%%%%%%%%%%%%%%%%%%%%%%%
\subsection{Methodology}
%%%%%%%%%%%%%%%%%%%%%%%%%%%%%%%%%%%%%%%%%%%%%%%%%%%%%
We start by writing the forced linear governing equations, here the compressible Navier-Stokes equations, as an input-output system in the frequency domain 
\begin{eqnarray}\label{eqn:inout}
\left (-\mathrm{i}\omega\mathsfbi{I}-\mathsfbi{A}_m\right )\qfft &=& \mathsfbi{B}\ffft, \\ \label{eqn:observable}
\yfft &=& \mathsfbi{C}\qfft,
\end{eqnarray}
where $\mathsfbi{A}_m$ is the linearized compressible Navier-Stokes operator, $\qfft$ is the state vector as before, and $\ffft$ the (for now unspecified) forcing. Equation (\ref{eqn:observable}) defines the output, or response, $\yfft$ as the product of an output matrix $\mathsfbi{C}$ with the state. Analogously, a input matrix $\mathsfbi{B}$ is introduced in (\ref{eqn:inout}). Without forcing, the right hand side of equation (\ref{eqn:inout}) is zero, and the global linear stability eigenvalue problem for $\qfft$ and $\omega$ is recovered. We use the same discretization scheme as in \citet{SchmidtEtAl_2017_JFM} and refer to that paper for details on the numerical method. 

Equations (\ref{eqn:inout}) allow us to express a direct relation between inputs and outputs,
\begin{equation}\label{eqn:yHf}
\yfft  = \simpleresolvent\ffft,
\end{equation}
by defining the resolvent operator
\begin{equation}\label{eqn:H}
\simpleresolvent=\mathsfbi{C}\left (-\mathrm{i}\omega\mathsfbi{I}-\mathsfbi{A}_m\right )^{-1}\mathsfbi{B}
\end{equation}
as the transfer function between them. We further define the modified, or weighted, resolvent operator
\begin{equation}
\label{eqn:resolventoperator} \resolvent = \mathsfbi{W_y}^{\frac{1}{2}}\simpleresolvent\mathsfbi{W_f}^{-\frac{1}{2}}=\tilde{\mathsfbi{Y}}\boldsymbol{\Sigma}\tilde{\mathsfbi{F}}^*
\end{equation}
that account for arbitrary inner products on the input and output spaces that are defined shortly. In the last equality of equation (\ref{eqn:resolventoperator}), we anticipated the result that the optimal responses $\tilde{\mathsfbi{Y}}=[\yffthat^{(1)}\;\yffthat^{(2)}\cdots\yffthat^{(N)}]$, forcings $\tilde{\mathsfbi{F}}=[\fffthat^{(1)}\;\fffthat^{(2)}\cdots\fffthat^{(N)}]$, and amplidude gains $\boldsymbol{\Sigma}=\text{diag}\left(\sigma_1,\sigma_2,\cdots,\sigma_N\right)$ can be found from the singular value decomposition (SVD) of the modified resolvent operator. By $\tilde{(\cdot)}$ we denote singular or eigenvectors. The modified resolvent operator in equation (\ref{eqn:resolventoperator}) is weighted such that the orthonormality properties 
\begin{eqnarray}
\left<\yffthat\ith,\yffthat\jth\right>_y&=&\yffthat\ithH\mathsfbi{W_y}\yffthat\jth = \delta_{ij}\text{ and}\\
\left<\fffthat\ith,\fffthat\jth\right>_f&=&\fffthat\ithH\mathsfbi{W_f}\fffthat\jth  = \delta_{ij}
\end{eqnarray}
hold for the optimal responses in the output norm $\left<\cdot,\cdot\right>_y$, and the forcings in the input norm $\left<\cdot,\cdot\right>_f$, respectively. The optimal forcings and responses are found from the definition of the optimal energetic gain
\begin{equation}
G^2_{max}(\ffft) = \max_{\|\ffft\|^2_f=1}\frac{\big<\yfft,\yfft\big>_y}{\big<\ffft,\ffft\big>_f} = \frac{\big<\yffthat,\yffthat\big>_y}{\big<\fffthat,\fffthat\big>_f} = G^2(\fffthat^{(1)}) = \sigma_1^{2}
\end{equation}
between inputs and outputs \citep[see \eg][]{schmid2001stability}. By writing the energetic gain in form of a generalized Rayleigh quotient and inserting equations (\ref{eqn:yHf}) and (\ref{eqn:H}), it is found that the orthogonal basis of forcings optimally ranked by energetic gain can be found from the eigenvalue problem 
\begin{equation}\label{evp2}
\mathsfbi{W_f}^{-1}    \mathsfbi{B}^H\left (-\mathrm{i}\omega\mathsfbi{I}-\mathsfbi{A}_m\right )^{-H}\mathsfbi{C}^H      \mathsfbi{W_y}\mathsfbi{C}\left (-\mathrm{i}\omega\mathsfbi{I}-\mathsfbi{A}_m\right )^{-1}\mathsfbi{B}\fffthat\ith = \sigma^2_i\fffthat\ith.
\end{equation}
Equation (\ref{evp2}) is solved by first factoring $\left(-\mathrm{i}\omega\mathsfbi{I}-\mathsfbi{A}_m\right )=\mathsfbi{L}\mathsfbi{U}$ \citep{sipp2012characterization} and then solving the eigenvalue problem
\begin{equation}\label{evp2}
\mathsfbi{W_f}^{-1}    \mathsfbi{B}^H \mathsfbi{U}^{-H}\mathsfbi{L}^{-H}\mathsfbi{C}^H      \mathsfbi{W_y}\mathsfbi{C}  \mathsfbi{U}^{-1}\mathsfbi{L}^{-1}  \mathsfbi{B}\tilde{\mathsfbi{F}} = \tilde{\mathsfbi{F}}\boldsymbol{\Sigma}^2
\end{equation}
for the largest eigenvalues using a standard Arnoldi method. The corresponding responses are readily obtained as
\begin{equation}\label{evp2}
\tilde{\mathsfbi{Y}} = \resolvent \tilde{\mathsfbi{F}} =\mathsfbi{W_y}^{\frac{1}{2}}\mathsfbi{C}\mathsfbi{U}^{-1}\mathsfbi{L}^{-1}\mathsfbi{B}\mathsfbi{W_f}^{-\frac{1}{2}}\tilde{\mathsfbi{F}}.
\end{equation}
In this study, we quantify both the energy of the input and the output in the compressible energy norm, as defined in equation (\ref{eqn:discnorm}), by setting $\mathsfbi{W_y}=\mathsfbi{W_f}=\mathsfbi{W}$. The matrices $\mathsfbi{B}$ and $\mathsfbi{C}$ we use for the sole purpose of restricting the analysis to the physical part of the computational domain by assigning zero weights to the sponge region. The physical domain corresponds to the domain of the LES data detailed in table \ref{tab:SPOD}. It is surrounded by a sponge region of width $D$, and discretized using $950\times195$ points in the streamwise and radial direction, respectively. The streamwise extent of the domain sets a limit of $\St\gtrsim0.2$ on the lowest possible frequency. For lower frequencies, the response structures become so elongated that domain truncation affects the gain. An upper limit on $\St$ is imposed by the numerical discretization, i.e.~the capability of the differentiation scheme to resolve the smallest structures in the response and forcing fields for the given resolution. This limit is $\St\lesssim2.5$ for the subsonic, $\St\lesssim1.5$ for the transsonic, and $\St\lesssim1$ for the supersonic case, respectively. A molecular Reynolds number of $\Rey=3\cdot10^4$ is used for this study. We return to this issue later and discuss our rationale in \S \ref{subsec:Re}.

%%%%%%%%%%%%%%%%%%%%%%%%%%%%%%%%%%%%%%%%%%%%%%%%%%%%%
\subsection{Resolvent spectra and modes} 
%%%%%%%%%%%%%%%%%%%%%%%%%%%%%%%%%%%%%%%%%%%%%%%%%%%%%

\begin{figure}
	\centerline{\includegraphics[trim=0 9.3cm 0 5mm, clip, width=1\textwidth]{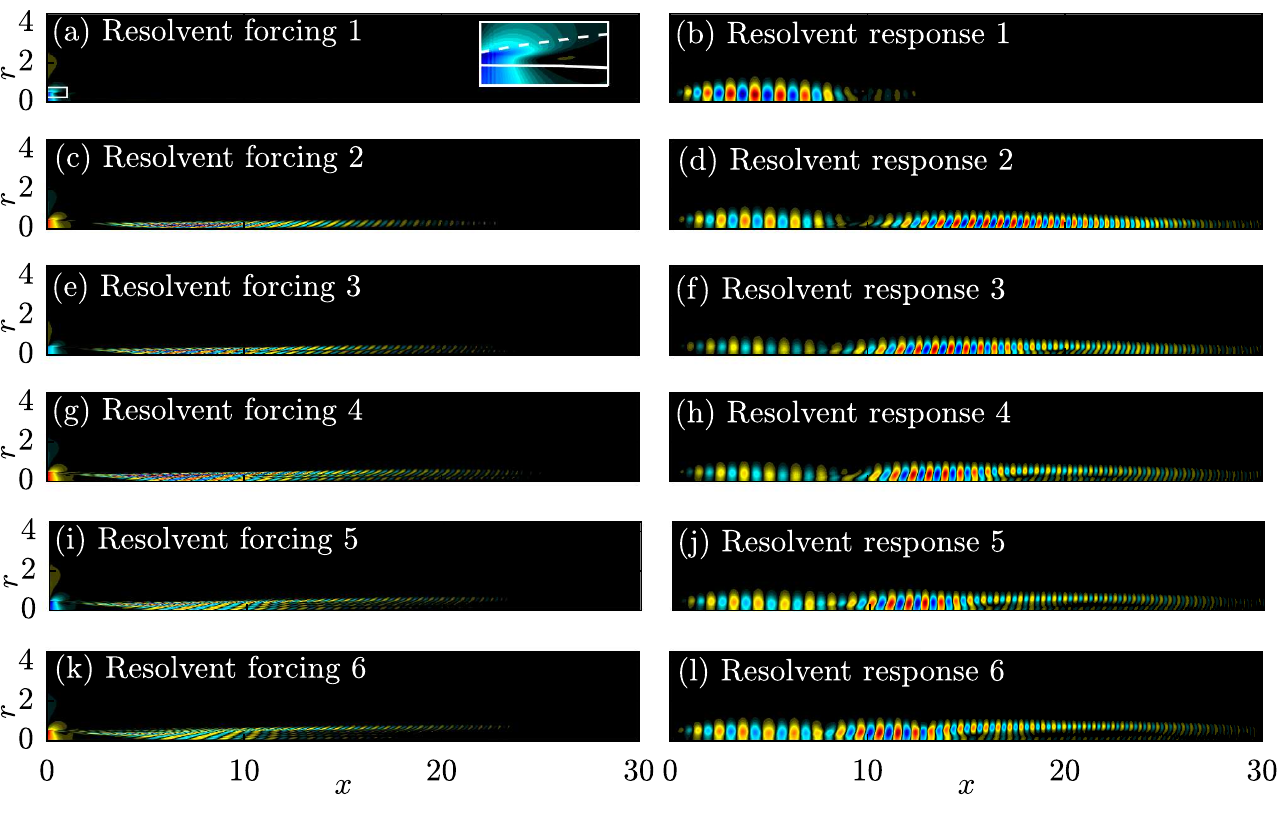}}% Images in 100% size
	\captionsetup{singlelinecheck=off}
	\caption[]{Optimal and suboptimal resolvent forcings (left) and corresponding responses (right) of the subsonic jet for $St=0.6$ and $m=0$. The pressure field (${\color{Red}\blacksquare}\!{\color{Black}\blacksquare}\!{\color{Blue}\blacksquare}$) is normalized with respect to its maximum absolute value. The optimal response mode in (a) is of KH type, whereas all suboptimal modes combine the KH and Orr-type waves. The inset in panel (a) shows the forcing structure close to the nozzle. The shear-layer and the potential core are outlined as in figure \ref{fig:M04_LESinst}.
}
	\label{fig:M04_Resolvent_St06}
\end{figure}
Figure \ref{fig:M04_Resolvent_St06} shows the optimal and the five leading suboptimal forcings and responses for $St=0.6$ and $m=0$. The leading mode in figure \ref{fig:M04_Resolvent_St06}(b) resembles a KH-type instability wavepacket that is confined to the initial shear-layer region. The corresponding optimal forcing in figure \ref{fig:M04_Resolvent_St06}(a) is confined close to the nozzle. The insert in figure \ref{fig:M04_Resolvent_St06}(a) reveals that the KH-type wavepacket is most efficiently forced by a structure that is oriented against the mean-shear in the vicinity of the lip-line. This is a typical manifestation of the Orr-mechanism and has similarly been observed in resolvent models of other flows \citep[\eg in][]{garnaud2013pref,dergham2013stochastic,jeun2016input,lesshafftmodeling,tissot2017}. The suboptimal modes in figure \ref{fig:M04_Resolvent_St06}(d,f,h,j,l) contain two wavepackets: one in the initial shear layer region that is similar to the KH wavepackets in the optimal mode and a second further downstream in the developing jet region. With increasing mode number, the downstream wavepacket moves upstream and becomes more spatially confined. It is optimally forced downstream of the inlet and over an axial distance comparable to the length of the response. Following the same arguments as for the suboptimal SPOD modes presented in figure \ref{fig:M04_SPODmodes_m013}, we term the downstream wavepackets Orr-type wavepackets. Both the KH-type and the Orr-type wavepackets are optimally exerted via the Orr-mechanism. From a local stability theory point of view, the two mechanisms are distinguished by their modal and non-modal nature, see \citet{jordan2017modal} and \citet{tissot2017}).

\begin{figure}
	\centerline{\includegraphics[trim=0 3.9cm 0 24mm, clip, width=1\textwidth]{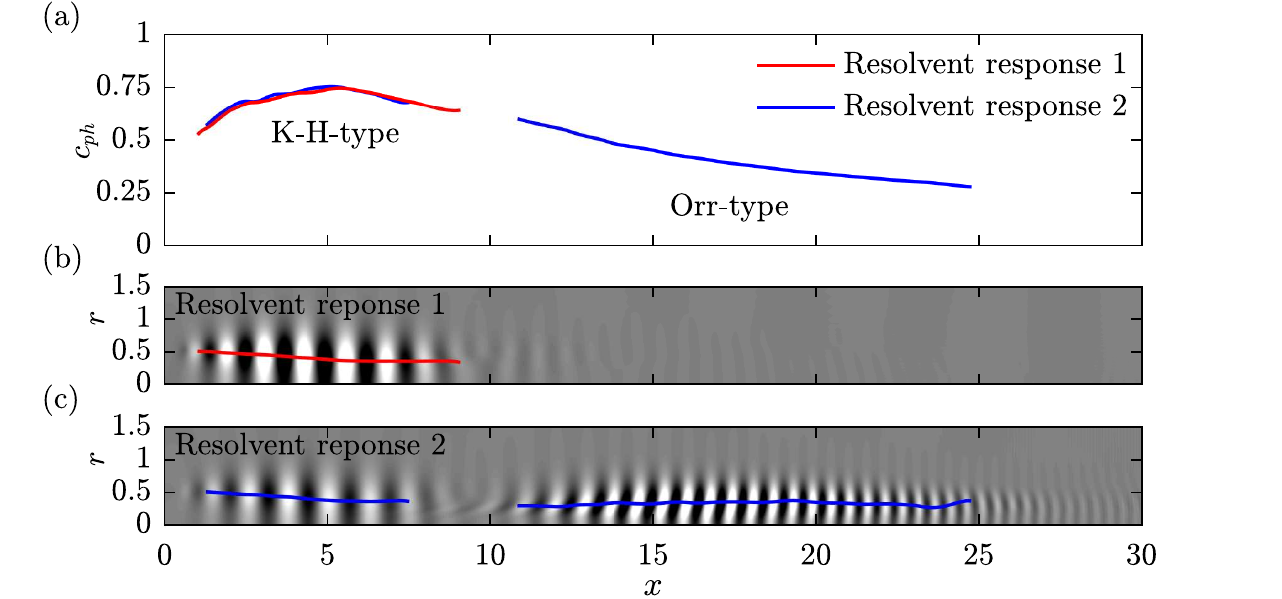}}% Images in 100% size
	\captionsetup{singlelinecheck=off}
	\caption[]{The critical layer effect in the subsonic jet for $St=0.6$ and $m=0$: 
	(a) estimated phase velocity (${\color{Red}\solidrule}$ first mode; ${\color{blue}\solidrule}$ second mode);	
	(b,c) normalized pressure of the first and second mode ($\blacksquare\!{\color{Gray}\blacksquare}\!\square$), and critical layer location where $c_{ph}=\bar{u}_x$. The KH and Orr wavepackets clearly follow the critical layer.
}
	\label{fig:M04_resolvent_phaseSpeed_St06}
\end{figure}
\citet{tissot2017} found that the critical layer, defined where the phase speed of the wavepacket is equal to the local mean velocity, plays an important role in the forced linear dynamics of jets. In accordance with our interpretation, their results suggests that the Orr-mechanism is active downstream of the potential core. Figure \ref{fig:M04_resolvent_phaseSpeed_St06}(a) shows the phase velocity of the leading and first suboptimal resolvent response for $m=0$ and $\St=0.6$. The phase velocity is approximated as $c_{ph}\approx \omega/\frac{\partial \theta_p}{\partial x}$, where $\theta_p=\arg(\phi_p)$ is the local phase of the pressure along $r=0.5$. The phase velocity is plotted for the regions where the pressure exceeds 25\% of its global absolute maximum value. The phase velocity of the upstream wavepacket in the initial shear-layer is almost identical for the first and second response mode. Besides their similar structure, this also suggests that the primary wavepacket of the suboptimal mode is of KH-type. The phase velocity of the downstream wavepacket decreases with axial distance in accordance with the jet's velocity-decay rate. Panels \ref{fig:M04_resolvent_phaseSpeed_St06}(b,c) show that the first and second wavepackets closely follow the critical layer. The variation of the phase speed with axial distance explains why the KH and Orr wavepackets are characterized by broad bands in frequency-wavenumber space, see figure \ref{fig:M04_StvsAlpha_SPOD}.

\begin{figure}
	\centerline{\includegraphics[trim=0 0cm 0 0mm, clip, width=1\textwidth]{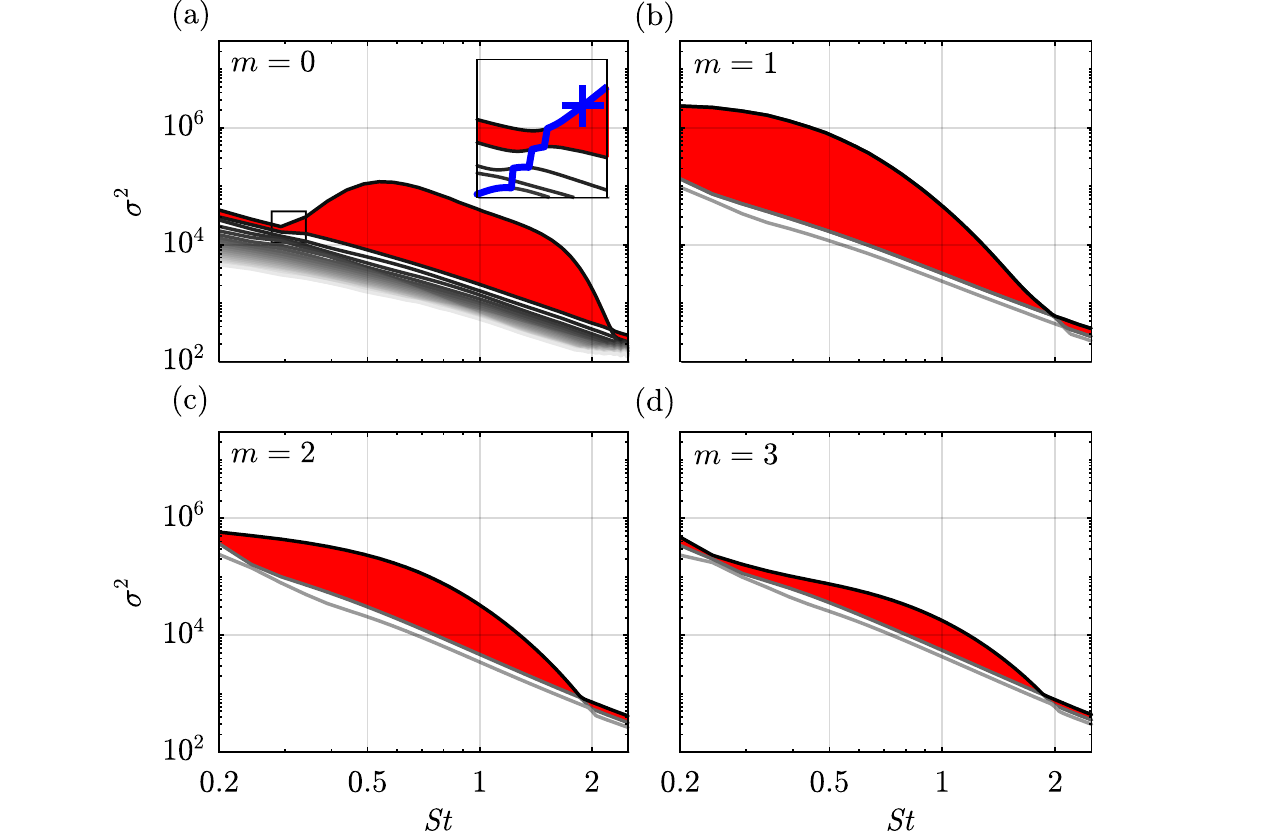}}% Images in 100% size
	\captionsetup{singlelinecheck=off}
	\caption[]{Optimal energetic gain spectra ($\blacksquare\!{\color{Gray}\blacksquare}\!\square$, $\sigma_1\!>\!\sigma_2\!>\!\cdots\!>\!\sigma_{N}$) for the subsonic jet: (a) $m=0$; (b) $m=1$; (c) $m=2$; (d) $m=3$. The difference between the optimal and the first suboptimal mode (${\color{Red}\blacksquare}$) highlights the low-rank behavior. The thirty largest singular values were computed for $m=0$, and the three largest for $m>0$. In the inset in panel (a), the KH-type mode is tracked (${\color{blue}\solidrule}$) into the frequency range where it becomes suboptimal. The leading mode at $\St=0.33$ (${\color{blue}\boldsymbol{+}}$) serves reference and the scalar projection $\langle\qvec^{(i)},\qvec^{(1)}_{\text{ref}}\rangle_E$ onto that mode is used for the tracking.}
	\label{fig:M04_gainVsSt_m0123}
\end{figure}
Resolvent gain spectra for the first four azimuthal wavenumbers are shown figure \ref{fig:M04_gainVsSt_m0123}. As in figure \ref{fig:SPODvsSPODforcing_spectrum_m01}, we highlight the difference between the leading and the first suboptimal gain to emphasize low-rank behavior. A pronounced low-rank behavior is evident for $0.3\lesssim\St\lesssim2$ for $m=0$ as can be seen in panel \ref{fig:M04_gainVsSt_m0123}(a). The 20 leading modes were calculated for $m=0$, and the the three leading modes for $m>0$. The inset in panel \ref{fig:M04_gainVsSt_m0123}(a) shows that the KH mechanism persists into lower frequencies as a suboptimal mode (${\color{blue}\solidrule}$). The reference mode is of pure KH-type, and it was confirmed by visual inspection of the mode shapes (not shown) that the KH signature indeed prevails in the suboptimal modes. The vertical line segments indicate where transitions to the next lower singular suboptimal. At high frequencies, the continuation is more obvious. For all four azimuthal wavenumbers, the curve associated with the KH wavepacket crosses the suboptimal gain curves at $\St\approx2$. This marks the end of the low-rank frequency band. For $m>0$, the low-rank band extends to lower frequencies. Panel \ref{fig:M04_gainVsSt_m0123}(b) shows that a strong low-rank behavior is predicted for frequencies even lower that the minimum frequency $\St<0.2$ for $m=1$. With increasing azimuthal wavenumber, the low-rank behavior becomes less and less pronounced. All these qualitative trends are reflected in the empirical SPOD energy spectra in figure \ref{fig:SPODvsSPODforcing_spectrum_m01}. The direct comparison of the SOPD analysis with the resolvent model is the subject of \S \ref{mach}. 

%%%%%%%%%%%%%%%%%%%%%%%%%%%%%%%%%%%%%%%%%%%%%%%%%%%%%
\subsection{Reynolds number effects}\label{subsec:Re}
%%%%%%%%%%%%%%%%%%%%%%%%%%%%%%%%%%%%%%%%%%%%%%%%%%%%%
\begin{figure}
	\centerline{\includegraphics[trim=0 0.5cm 0 0mm, clip, width=1\textwidth]{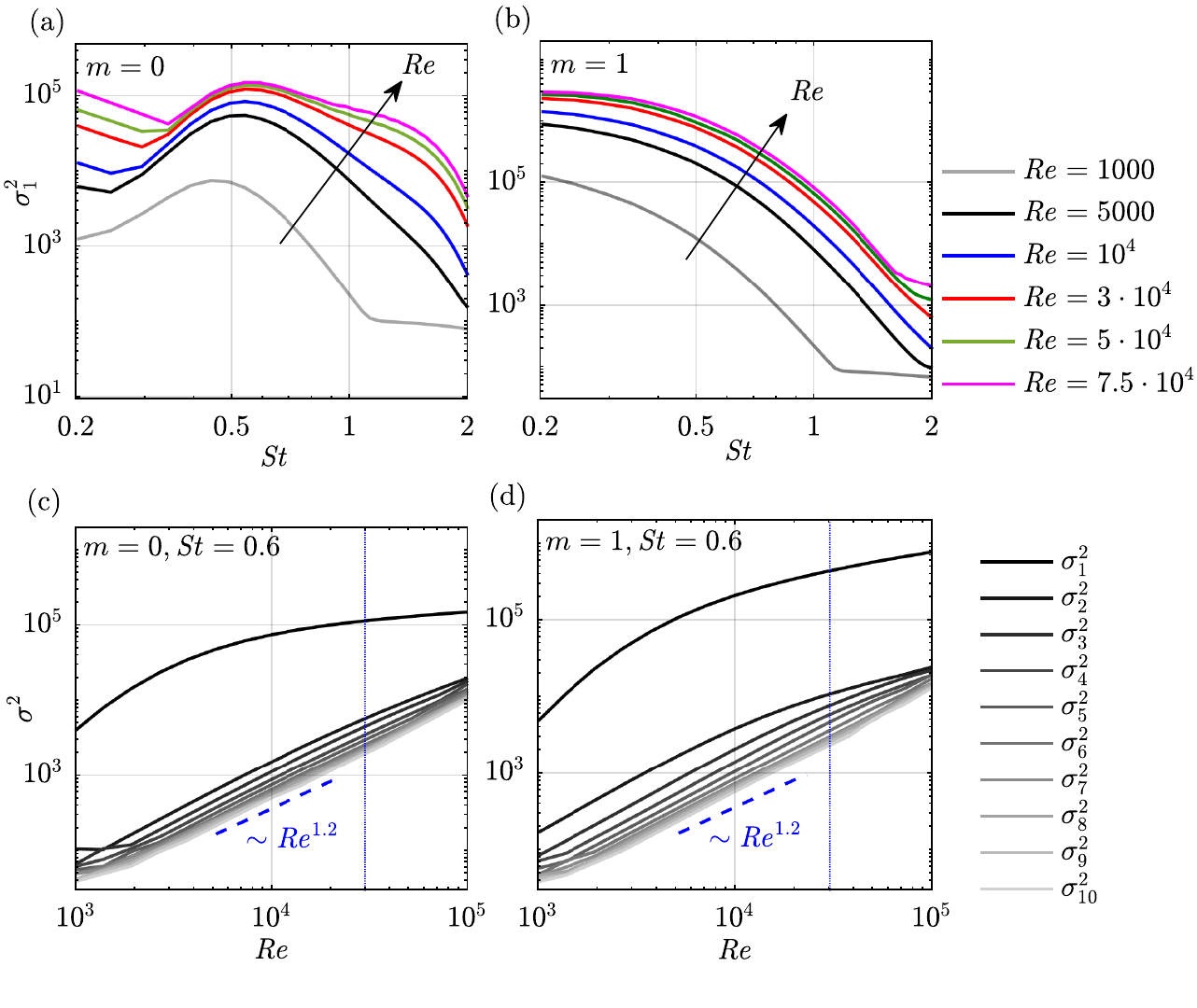}}% Images in 100% size
	\captionsetup{singlelinecheck=off}
	\caption[]{Reynolds number effect on the energetic gain of the subsonic jet: (a,b) optimal gain spectra for $m=0$ and $m=1$; (c,d) energetic gain at $\St=0.6$ for $m=0$ and $m=1$. The blue dotted line (${\color{blue}\cdots}$) marks the Reynolds number $\Rey=3\cdot10^4$ used for the present study.} 
	\label{fig:M04_resolvent_ReStudy}
\end{figure}
The effect of the Reynolds number on the spectrum of the discretized linearized Navier-Stokes operator $\mathsfbi{A}_m$ is studied in \citet[appendix D]{SchmidtEtAl_2017_JFM}. In figure \ref{fig:M04_resolvent_ReStudy}, the effect of the Reynolds number on the resolvent gain is investigated. The optimal gain $\sigma_1^2$ for $m=0$ and $m=1$ shown in panels \ref{fig:M04_resolvent_ReStudy}(a) and \ref{fig:M04_resolvent_ReStudy}(b), respectively, increases with increasing Reynolds number over the entire frequency range. Above $\Rey\geq3\cdot10^4$, the gain does not show a pronounced dependency on the Reynolds number for all but the lowest frequencies for $m=0$. The gain of the leading ten modes at a fixed frequency of $\St=0.6$ is shown below in \ref{fig:M04_resolvent_ReStudy}(c,d). At this particular frequency, the optimal gain is significantly higher that the suboptimal gains, which are all of comparable magnitude. Similar to the modal energy gap in the SPOD analysis, this reflects the low-rank behavior of the jet in the resolvent model. The suboptimal gains follow an approximate power law behavior over the range of Reynolds number studied. The different behavior of the optimal and the suboptimal gains highlights the disparate physical nature of the two mechanisms. Their different scalings highlight the importance of a proper choice of Reynolds number for mean-flow-based resolvent models. In particular, the model Reynolds number determines which of the two mechanisms dominates at a certain frequency. This becomes apparent, for example, for $m=0$ at low frequencies: the kink in the gain curves in figure \ref{fig:M04_resolvent_ReStudy}(a), which marks the transition from one scaling to another, shifts to higher frequencies with increasing Reynolds number. For $\Rey=7.5\cdot10^4$, the exchange of optimal mechanisms occurs at $\St\approx0.35$, whereas it occurs at $\St\approx0.25$ for $\Rey=7.5\cdot10^4$. We chose a Reynolds number of $\Rey=3\cdot10^4$ for the present study. This choice is motivated by the good correspondence with the LES, as discussed in the next section. For now, and in absence of a proper model for the effective Reynolds number, the Reynolds number has to be understood as a free model parameter. \citet{mettot2014quasi}, for example, demonstrate that resolvent analyses based on the linearized Reynolds-averaged Navier-Stokes (RANS) equations with modeled turbulence do not necessarily give superior results to our ad-hoc approach. 

%%%%%%%%%%%%%%%%%%%%%%%%%%%%%%%%%%%%%%%%%%%%%%%%%%%%%%%%%%%%%%%%%%%%%%%%%%%%%%%%%%%%
\section{Comparison of the SPOD and resolvent models}\label{comparison}
%%%%%%%%%%%%%%%%%%%%%%%%%%%%%%%%%%%%%%%%%%%%%%%%%%%%%%%%%%%%%%%%%%%%%%%%%%%%%%%%%%%%
In this section, we make comparisons between the high-energy SPOD modes and the high-gain resolvent modes.  This comparison is facilitated by recently established theoretical connections between the two methods \citep{towne2015stochastic,lesshafftmodeling,TowneEtAt2017SPOD}.  Specifically, the resolvent operator relates the cross-spectral density of the nonlinear forcing to the cross-spectral density of the response.  IF the forcing were uncorrelated in space and time with equal amplitude everywhere, i.e., unit-variance white noise, then the SPOD and resolvent modes would be identical.  This result is conceptually intuitive--when there is no bias in the forcing, the modes with highest gain are also the most energetic.

The nonlinear forcing terms in real turbulent flows are, of course, not white. \cite{ZareEtAl_2017_JFM} showed that it is necessary to account for correlated forcing in order to reconstruct the flow statistics of a turbulent channel flow using a linear model.  Of particular relevance to our study, \cite{TowneEtAl_2017_AIAA} investigated the statistical properties of the nonlinear forcing terms in a turbulent jet.  They found limited correlation in the near-nozzle shear-layer but significant correlation further downstream, especially near and beyond the end of the potential core. 

Correlated nonlinear forcing leads to differences between SPOD and resolvent modes.  Precisely, the correlation causes a bias in the forcing that preferentially excites certain resolvent modes and mixes them together via correlations between different modes \citep{TowneEtAt2017SPOD}.  As a result, multiple resolvent modes are required to reconstruct each SPOD mode.  Accordingly, we do not expect a one-to-one correspondence between the SPOD and resolvent modes of the jet.  Rather, we are looking for the signatures of the high-gain resolvent modes within the high-energy SPOD modes.  That is, we seek evidence that the mechanisms identified in the resolvent modes are active in the real flow and responsible for the most energetic coherent structures.

%Recently, a theoretical connection between SPOD and the resolvent operator has been made \citep{towne2015stochastic,lesshafftmodeling,TowneEtAt2017SPOD}: SPOD and resolvent modes are identical in the case of white-noise forcing. As discussed in \S \ref{resolvent}, we are interested in the forced linear dynamics of the jet, where the forcing is a result of nonlinear interactions within the flow. While the nonlinear terms in real flows, including turbulent jets, are not white \citep{towne2015stochastic,ZareEtAl_2017_JFM,TowneEtAl_2017_AIAA}, this theoretical connection is helpful for interpreting the following comparisons between SPOD and resolvent.

\begin{figure}
	\centerline{\includegraphics[trim=0 0cm 0 1mm, clip, width=1\textwidth]{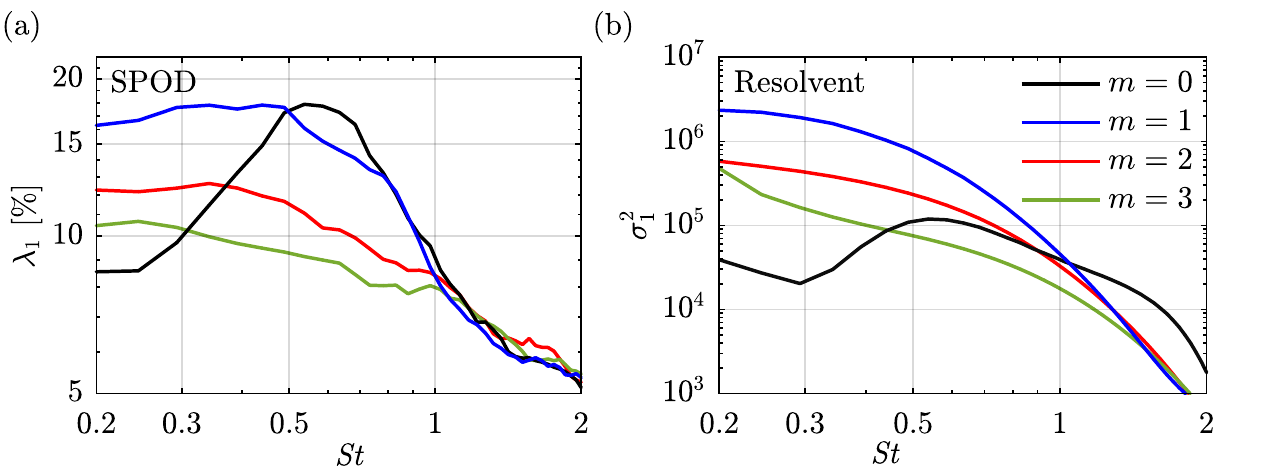}}% Images in 100% size
	\captionsetup{singlelinecheck=off}
	\caption[]{First SPOD mode energy spectra (a) and optimal resolvent gain (b) for $0\leq m \leq 3$ for the subsonic jet. A five-point moving average filter was used to increase the clarity of the SPOD spectrum.}
	\label{fig:M04_resolventVsSPOD_spectrum_mStudy_globalForcing}
\end{figure}
The leading mode SPOD energy spectra (see figure \ref{fig:SPODvsSPODforcing_spectrum_m01}) and optimal resolvent gain curves (see figure \ref{fig:M04_gainVsSt_m0123}) for $m=0,\cdots,3$ are compared in figure \ref{fig:M04_resolventVsSPOD_spectrum_mStudy_globalForcing}. In panel \ref{fig:M04_resolventVsSPOD_spectrum_mStudy_globalForcing}(a), we show the relative energy of the leading mode in percentages of the total energy at each frequency. We choose this quantity as a qualitative surrogate for the gain, which is not defined for the LES data. The resolvent gain curves capture the trends of the SPOD eigen-spectra remarkably well. The peak in relative energy of the leading $m=0$ SPOD mode in figure \ref{fig:M04_resolventVsSPOD_spectrum_mStudy_globalForcing}(a) clearly indicates low-rank behavior. At low frequencies, the ordering of the resolvent gains is directly reflected in the relative importance of the corresponding SPOD modes. 

\begin{figure}
	\centerline{\includegraphics[trim=0 9.1cm 0 4mm, clip, width=1\textwidth]{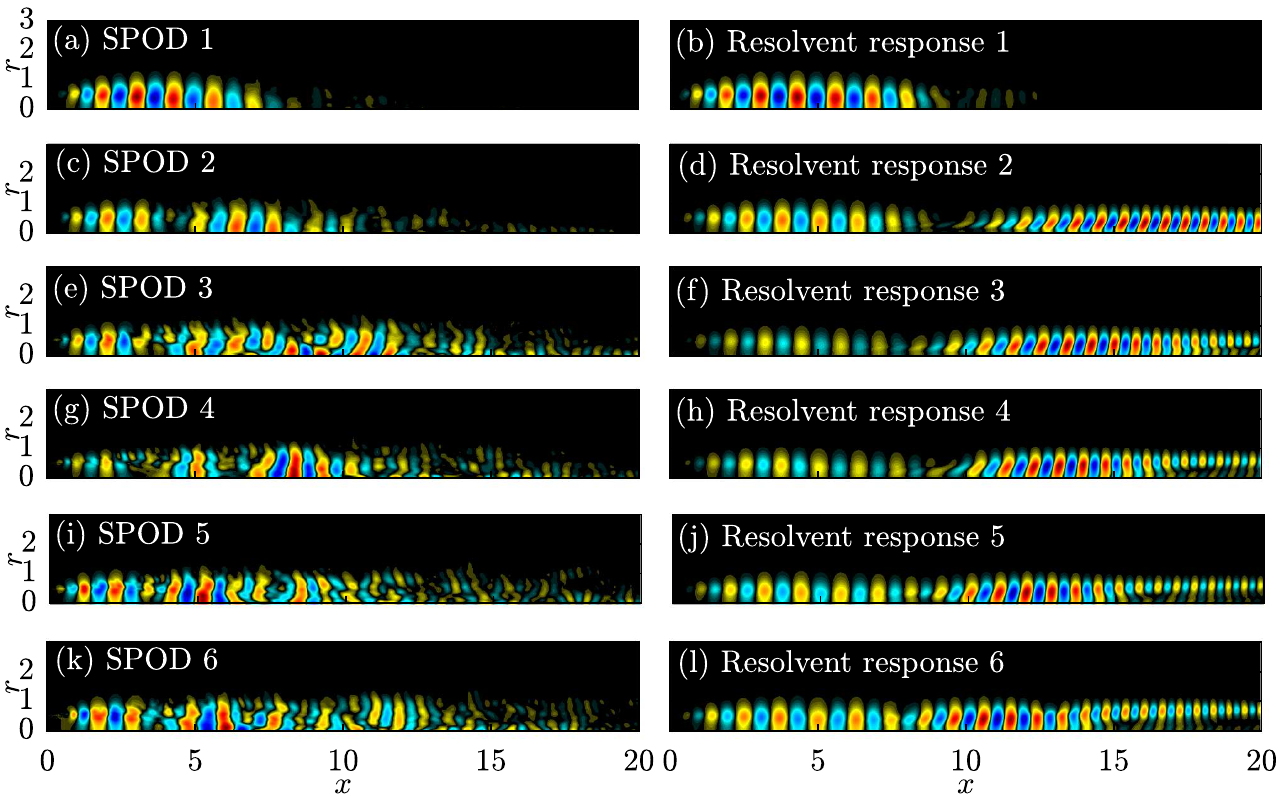}}% Images in 100% size
	\captionsetup{singlelinecheck=off}
	\caption[]{Empirical SPOD modes (left) and optimal resolvent response modes (right) for $St=0.6$ and $m=0$ for the subsonic jet. The normalized pressure is shown. The leading resolvent response in (b) accurately models leading SPOD mode in (a).}
	\label{fig:M04_SPODvsResolvent_St06}
\end{figure}
A comparison of the six leading resolvent and SPOD modes for $m=0$ and $\St=0.6$ is shown in figure \ref{fig:M04_SPODvsResolvent_St06}. The KH-type wavepacket of the first SPOD mode in \ref{fig:M04_SPODvsResolvent_St06}(a) closely resembles the optimal resolvent mode in figure \ref{fig:M04_SPODvsResolvent_St06}(b). Unlike the resolvent responses (as discussed in the context of figure \ref{fig:M04_Resolvent_St06}), the subdominant SPOD modes do not follow an immediately obvious hierarchy. Similar to the resolvent modes, they exhibit a multi-lobe structure and peak downstream or at the end of the potential core. Close to the nozzle, their structure resembles the KH-type waveform of the first mode. Their highly distorted structure suggests that their statistics may not be as well converged as the leading mode. The first three modes are characterized by an increasing integer number of lobes or successive wavepackets (this becomes much more evident in figure \ref{fig:M04_StvsX_SPODvsResolvent} below).

\begin{figure}
	\centerline{\includegraphics[trim=0 5.5cm 0 3mm, clip, width=1\textwidth]{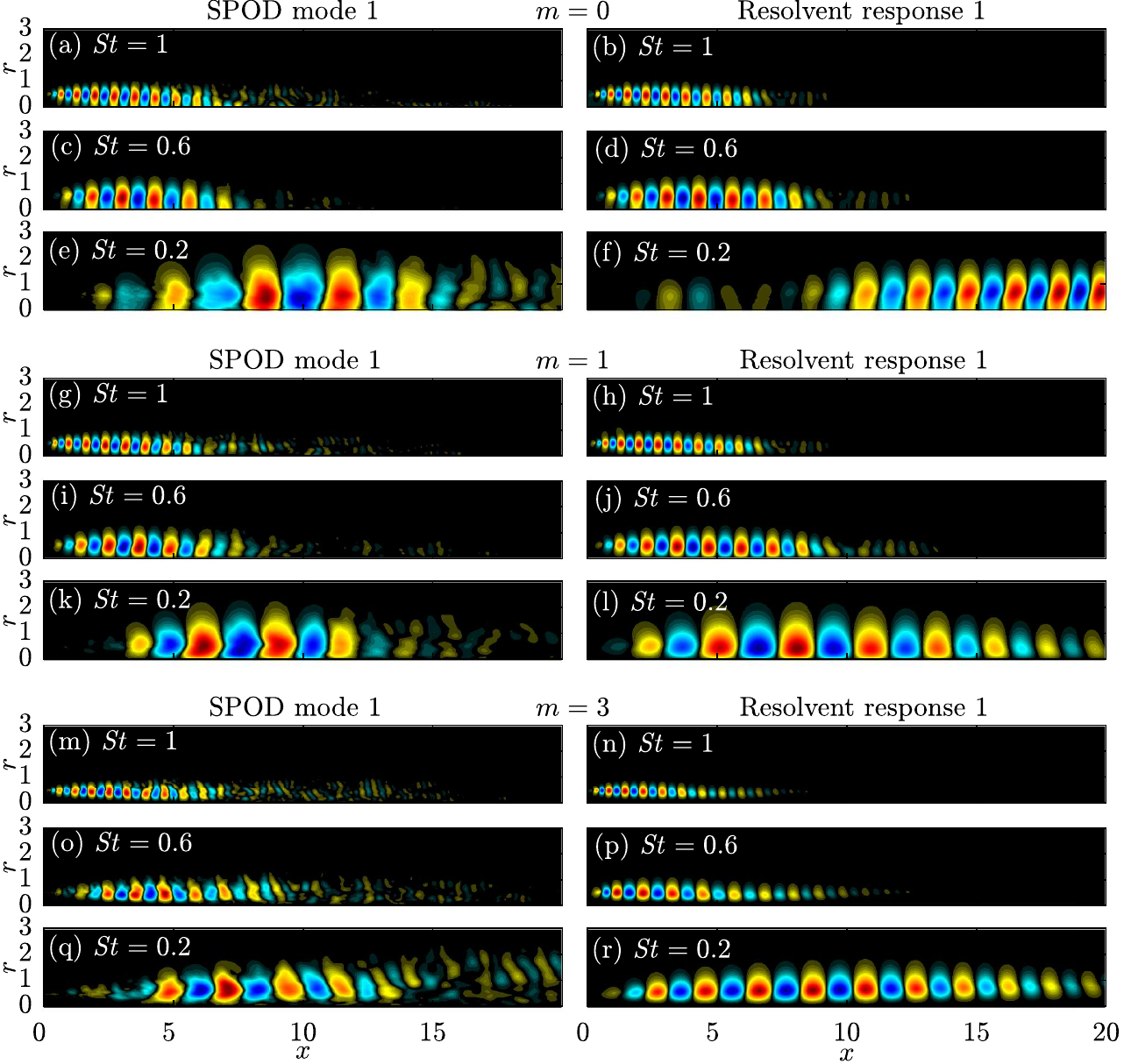}}% Images in 100% size
	\captionsetup{singlelinecheck=off}
	\caption[]{Comparison between leading empirical SPOD modes (left) and optimal resolvent response modes (right) at three representative frequencies for the subsonic jet: (a-f) $m=0$; (g-l) $m=1$; (m-r) $m=3$. The normalized pressure is shown. The leading modes compare favorably with the exception of the low frequency case for $m=0$ shown in (e,f), where the SPOD mode is of KH type, whereas the optimal response mode is of Orr type. Only part of the computational domain is shown for clarity.}
	\label{fig:M04_ResolventVsSPOD_m013}
\end{figure}
Figure \ref{fig:M04_ResolventVsSPOD_m013} makes direct comparisons between the leading SPOD and resolvent modes for different frequencies ($\St=1.0,0.6,0.2$) and azimuthal wavenumbers ($m=0,1,3$). In general, the modes compare well for frequency-azimuthal wavenumber combinations that exhibit low-rank behavior according to the SPOD and resolvent gain spectra in figures \ref{fig:SPODvsSPODforcing_spectrum_m01} and \ref{fig:M04_gainVsSt_m0123}, respectively. For example, good agreement is obtained for $\St=0.6$ and $\St=1.0$ for $m=0$ and $m=1$, i.e.~figure \ref{fig:M04_ResolventVsSPOD_m013}(a-d,g-j). In figure \ref{fig:M04_ResolventVsSPOD_m013}(e,f), the leading resolvent mode is an Orr-type wavepacket far downstream of the potential core, whereas the SPOD mode peaks further upstream and has a larger axial wavelength. For $m=3$ in figure \ref{fig:M04_ResolventVsSPOD_m013}(m-q) and higher azimuthal wavenumbers, the SPOD modes appear to be less well converged as compared to their low $m$ counterparts. This is, again, explained by the observation that the low-rank behavior decreases as $m$ increases.  

\begin{figure}
	\centerline{\includegraphics[trim=0 9.4cm 0 10mm, clip, width=1\textwidth]{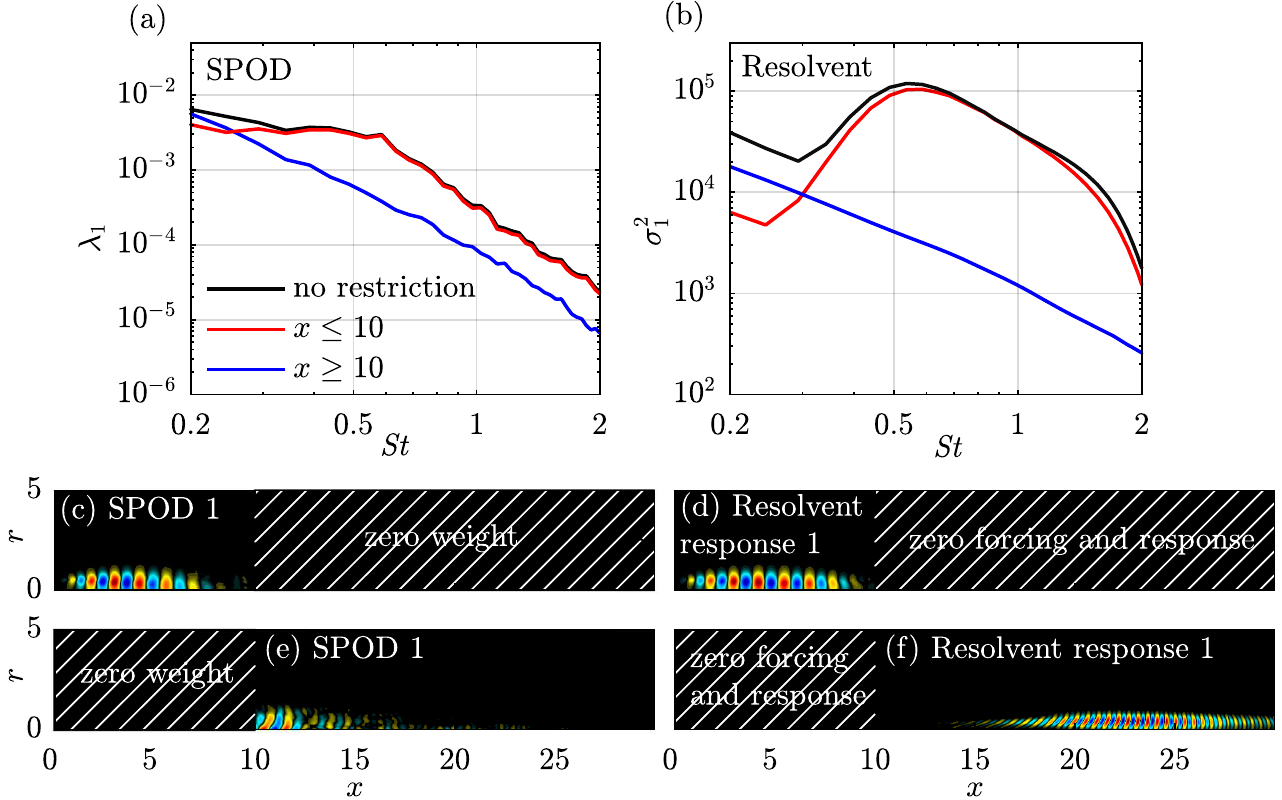}}% Images in 100% size
	\captionsetup{singlelinecheck=off}
	\caption[]{Isolation of the mechanisms for spatial modal growth in the initial shear layer, and non-modal spatial growth further downstream for the subsonic jet: (a) SPOD spectra; (b) optimal resolvent gain spectra; (c-f) normalized pressure field for $St=0.6$. For the SPOD, the restriction is realized via the weight matrix and the non-zero weighted regions are shown. For the resolvent analysis, the forcing and response are restricted via the input and output matrices.}
	\label{fig:M04_resolventAndSPOD_spectra_fLeftRightGlobal_m0}
\end{figure}

Both the SPOD and resolvent methodologies allow us to isolate the characteristics of the KH-type wavepackets near the nozzle and the downstream Orr-type wavepackets through their different spatial support. In the SPOD analysis, we utilize the weight matrix $\mathsfbi{W}$ to assign zero weight to the region we wish to exclude, e.g. $x>10$, to focus on the initial shear-layer region and vice versa for the developing jet region. The resulting energy spectra and modes are depicted in figure \ref{fig:M04_resolventAndSPOD_spectra_fLeftRightGlobal_m0}(a,c,e). For the resolvent analysis, we restrict both the forcing and the response to the region of interest through the input and output matrices $\mathsfbi{B}$ and $\mathsfbi{C}$. The resulting gain spectra and modes are shown in figure \ref{fig:M04_resolventAndSPOD_spectra_fLeftRightGlobal_m0}(b,d,f). The SPOD and gain spectra consistently separate the two mechanisms. In both cases, the spectra obtained without spatial restriction are a superposition of the spectra of the two isolated physical mechanisms. The different spatial support of the restricted SPOD (panel \ref{fig:M04_resolventAndSPOD_spectra_fLeftRightGlobal_m0}(e)) and resolvent (panel \ref{fig:M04_resolventAndSPOD_spectra_fLeftRightGlobal_m0}(f)) modes emphasizes that the forcing of the subdominant mode is clearly not white, though there are obvious similarities in the wavepacket shape indicating a similar mechanism.

%The structural difference between the downstream-restricted SPOD and resolvent modes in panel \ref{fig:M04_resolventAndSPOD_spectra_fLeftRightGlobal_m0}(e) and \ref{fig:M04_resolventAndSPOD_spectra_fLeftRightGlobal_m0}(f), respectively, is explained by the enforced non-low-rank nature: the KH instability mechanism is mitigated through the weighting, which leaves only non-modal dynamics \hlcyan{[This last sentence is terrible. I think it has to go... I don't even know what I tried to say here; could you address in your forcing section why 13(e) and (f) cannot expected to look similar? Then we could refer to that section here.]}. 

\begin{figure}
	\centerline{\includegraphics[trim=0 0.3cm 0 4mm, clip, width=1\textwidth]{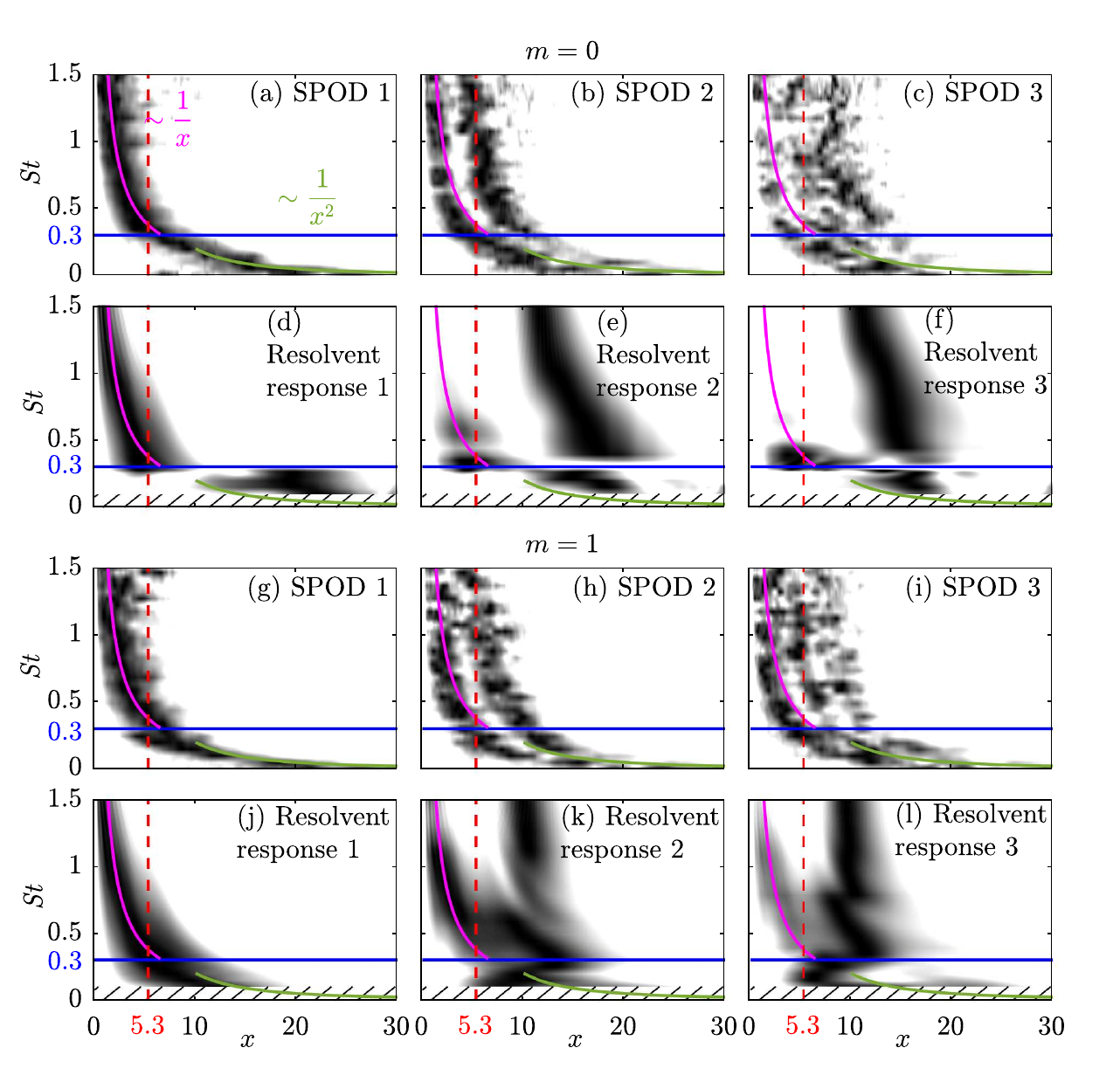}}% Images in 100% size
	\captionsetup{singlelinecheck=off}
	\caption[]{Frequency-axial distance diagrams ($\blacksquare\!{\color{Gray}\blacksquare}\!\square$,$\tilde{P}_{pp}$) along the lip-line ($r_0=0.5$) for $m=0$ (top half) and $m=1$ (bottom half) for the subsonic jet. The three leading SPOD and resolvent modes are compared. The estimated pressure PSD, $\bar{P}_{pp}$, is normalized by its maximum value at each frequency. The end of the potential core ({\color{red} {\dashedrule}}, $x=5.3$), and a Strouhal number of $\St=0.3$ ({\color{blue} {\solidrule}}) differentiate \emph{shear-layer} from \emph{developing jet} behavior. The different frequency scalings of shear-layer ({\color{Magenta} {\solidrule}}, $\St\sim\frac{1}{x}$) and jet wavepackets ({\color{Green} {\solidrule}}, $\St\sim\frac{1}{x^2}$) are indicated.}
	\label{fig:M04_StvsX_SPODvsResolvent}
\end{figure}
Plotting the temporal and azimuthal PSD 
\begin{equation}
\tilde{P}_{qq} = \frac{1}{n_{\text{blk}}}\sum_{l=1}^{n_{\text{blk}}} \left|\hat{q}^{(l)}_{m\omega}(x,r=r_0)\right|^2
\end{equation}	
as a function of $x$ along a line of constant distance $r_0$ from the axis allows us to locate the wavepackets in space. The resulting frequency-axial distance diagrams are shown in figure \ref{fig:M04_StvsX_SPODvsResolvent} for $m=0$ and $m=1$. The results for the SPOD and the resolvent response modes are directly compared. This form of visualization of the SPOD modes brings to light the hierarchal structure of the SPOD modes more clearly. From figure \ref{fig:M04_ResolventVsSPOD_m013}a-c for $m=0$, and \ref{fig:M04_ResolventVsSPOD_m013}g-i  for $m=1$, respectively, it becomes apparent that the higher order modes are characterized by an increasing number of subsequent wavepackets in the streamwise direction.

The change from KH-type shear-layer instability to Orr-type wavepackets in the developing jet is apparent from the change of the slope of the PSD at $(x,\St)\approx(6,0.3)$. Different frequency scalings in the two regions explain this sudden change. They can be directly deduced from the varying characteristic velocity and length scales of each region as shown in figure \ref{fig:baseFlow_ucl_r12}. The initial shear-layer grows linearly while the characteristic velocity stays constant. The frequency of the high-frequency wavepackets therefore scales with $1/x$. In the developing and self-similar jet regions, the jet width increases linearly, but the centerline velocity decays inversely proportional to the axial distance. The frequency in that region consequentl y scales with $1/x^2$. The different frequency-scalings in the two regions of the jet are apparent in other studies on wavepacket modeling \citep{CavalieriSasakiJordanEtAl2016,SasakiEtAl2017}, and on acoustic-source localization \citep{bishop1971noise,SchlinkerEtAl_2009_AIAA}.

The optimal response modes in figure \ref{fig:M04_StvsX_SPODvsResolvent}(d) and \ref{fig:M04_StvsX_SPODvsResolvent}(j) accurately predict the wavepacket location in the initial shear-layer region in the leading SPOD modes. For $m=1$ in figure \ref{fig:M04_StvsX_SPODvsResolvent}(j), the low-rank behavior of the jet permits accurate predictions at low frequencies. For $m=0$ (panel \ref{fig:M04_StvsX_SPODvsResolvent}(d)), on the contrary, the non-low-rank behavior at low frequencies hinders a rank-one resolvent-mode representation of the leading SPOD mode. Similarly, the subdominant SPOD modes shown in panels \ref{fig:M04_StvsX_SPODvsResolvent}(b,c) and \ref{fig:M04_StvsX_SPODvsResolvent}(h,i) cannot be represented by a single suboptimal resolvent mode.

\begin{figure}
	\centerline{\includegraphics[trim=0 6.4cm 0 2mm, clip, width=1\textwidth]{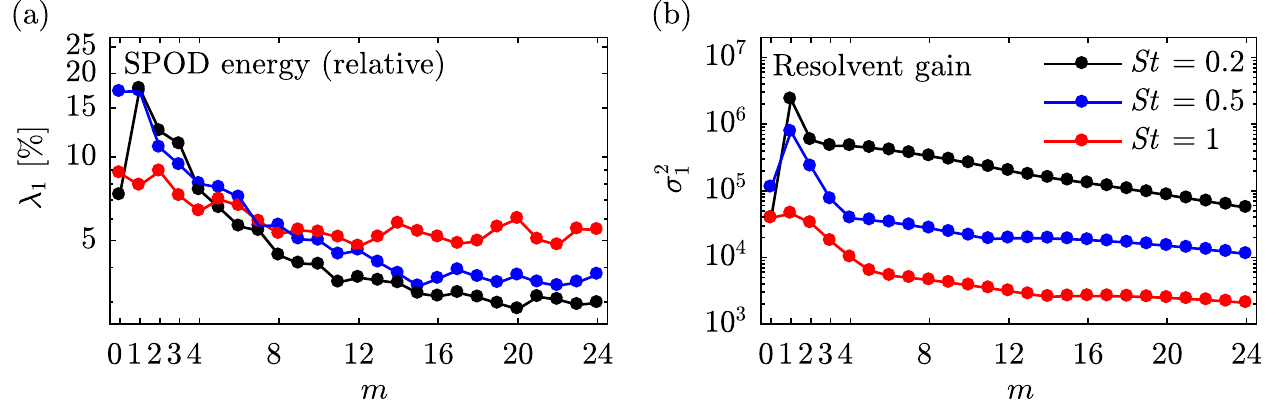}}% Images in 100% size
	\captionsetup{singlelinecheck=off}
	\caption[]{Azimuthal wavenumber dependence of the relative modal energy of the SPOD (a), and the optimal resolvent gain (b) for three representative frequencies fir the subsonic jet.}
	\label{fig:M04_resolventAndSPOD_vs_m_percent}
\end{figure}

The azimuthal wavenumber dependence of the SPOD energy and the resolvent gain is investigated in figure \ref{fig:M04_resolventAndSPOD_vs_m_percent}. As in figure, \ref{fig:M04_resolventVsSPOD_spectrum_mStudy_globalForcing}, we show the percentage of the energy of the first SPOD mode to highlight low-rank behavior. The falloff of the SPOD energy spectra seen in panel \ref{fig:M04_resolventAndSPOD_vs_m_percent}(a) implies that the low-rank behavior is more pronounced at low azimuthal wavenumbers and lower frequencies. For higher frequencies such as $\St=1$, the relative energy of the first SPOD mode is not a strong function of the azimuthal wavenumber. At higher azimuthal wavenumbers $m\gtrsim10$, its relative energy content stays at $\approx5\%$, which is above the levels of the two lower frequency cases. Similar trends are observed for the resolvent gain shown in panel \ref{fig:M04_resolventAndSPOD_vs_m_percent}(b). The maximum gain, for example, is attained for $m=1$, and the gain curve for the highest frequency is a much weaker function of the azimuthal wavenumber than in the case of two lower frequencies. For higher azimuthal wavenumbers, the gain falls off almost monotonically.

%%%%%%%%%%%%%%%%%%%%%%%%%%%%%%%%%%%%%%%%%%%%%%%%%%%%%%%%%%%%%%%%%%%%%%%%%%%%%%%%%%%%  
\section{Effect of the Mach number}\label{mach}
%%%%%%%%%%%%%%%%%%%%%%%%%%%%%%%%%%%%%%%%%%%%%%%%%%%%%%%%%%%%%%%%%%%%%%%%%%%%%%%%%%%%   
In the following, we addresses the effect of compressibility on the low-rank behavior. SPOD and resolvent analyses are conducted for the other two LES cases with $M_j=0.9$ and $M_j=1.5$, respectively (see table \ref{tab:LES}). The main conclusions drawn from the analysis of the subsonic case in \S\S \ref{SPOD}-\ref{resolvent} regarding the behavior of the KH- and Orr-type wavepackets apply to the other regimes as well. We therefore catalogue the complete results for the two additional higher Mach number cases in appendix \ref{app:mach}, and focus on specific Mach number dependent physical effects in this section.

\begin{figure}
	\centerline{\includegraphics[trim=0 7.5cm 0 2mm, clip, width=1\textwidth]{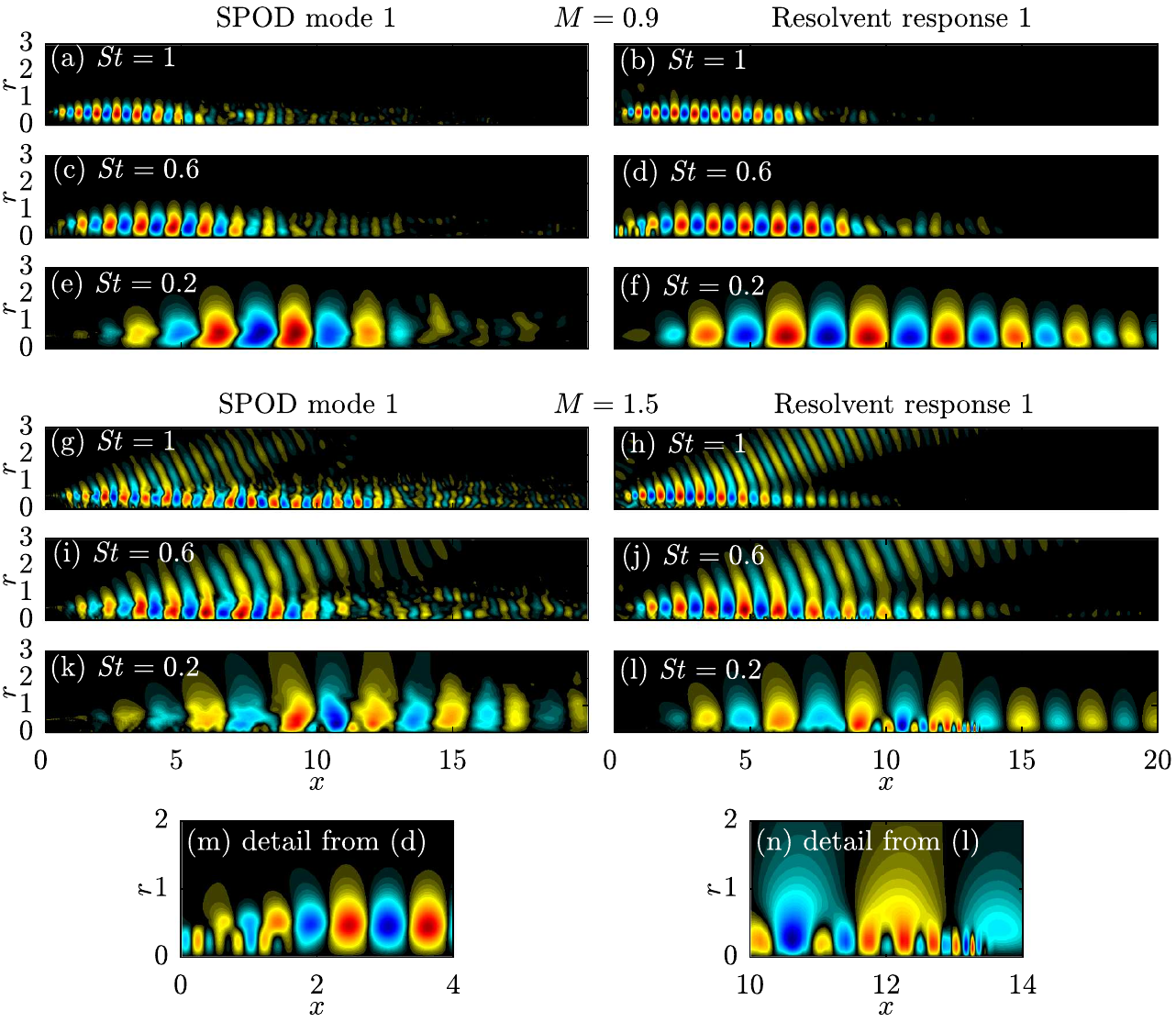}}% Images in 100% size
	\captionsetup{singlelinecheck=off}
	\caption[]{Comparison between leading empirical SPOD modes and optimal resolvent response modes at three representative frequencies for $m=1$ for the transsonic (a-f) and the supersonic jet (g-l). Panels (m) and (n) zoom in on a trapped acoustic mode \citep{TowneEtAl_2017_JFM,SchmidtEtAl_2017_JFM} in the transonic jet, and an upstream-propagating subsonic mode \citep{tam1989three} in the supersonic jet, respectively. The normalized pressure is shown.}
	\label{fig:M09vsB118_ResolventVsSPOD_m1}
\end{figure}  
Figure \ref{fig:M09vsB118_ResolventVsSPOD_m1} shows a side-by-side comparison of SPOD and resolvent modes for the transsonic and the supersonic jet. The leading modes are shown for three frequencies. 
Similar to the subsonic case in figure \ref{fig:M04_ResolventVsSPOD_m013}, favorable agreement between the empirical modes and the model is found. Significant discrepancies in terms of the length of the wavepackets and their radial structure is only observed for the transsonic jet at the lowest frequency, as shown in panels \ref{fig:M09vsB118_ResolventVsSPOD_m1}(e,f). In panels \ref{fig:M09vsB118_ResolventVsSPOD_m1}(g-j), it can be seen that the resolvent model accurately predicts the super-directive Mach wave radiation of the supersonic jet. 
\citet{sinha2014wavepacket} found similarly good agreement with their PSE model. 

Besides the KH and Orr-type wavepackets, which are vortical, jets also support different types of frequency-dependent acoustic waves. The transsonic jet, for example, supports trapped acoustic waves within the potential core \citep{TowneEtAl_2017_JFM,SchmidtEtAl_2017_JFM}. Such a trapped acoustic wave can be seen close to the nozzle in the detail shown in panel \ref{fig:M09vsB118_ResolventVsSPOD_m1}(m). In the supersonic jet, a closely related mechanism \citep{tam1989three,TowneEtAl_2017_JFM} is visible further downstream in panel \ref{fig:M09vsB118_ResolventVsSPOD_m1}(n). The reader is referred to \citep{TowneEtAl_2017_JFM} for details. In the present context, it suffices to recapitulate that the trapped waves are the result of an acoustic resonance in the transsonic jet regime $0.82<M_j<1$. More important than their physical nature for the present study is the observation that the resolvent analysis emphasized this type of intrinsic mechanism. This becomes clear from a closer inspection of figure \ref{fig:M09vsB118_ResolventVsSPOD_m1}(c,d) and \ref{fig:M09vsB118_ResolventVsSPOD_m1}(k,l), respectively. The acoustic wave phenomena are evident in the SPOD modes, but are more pronounced in the resolvent modes. Two factors contribute to this fact. First, the frequency and Reynolds number scaling of the  various physical effects is different, which leads to the same modeling challenges as for the KH and Orr-type modes discussed in the context of figure \ref{fig:M04_resolvent_ReStudy} (see also appendix \ref{app:mach}, figure \ref{fig:B118_gainVsSt_m0123}). Second, efficient means of forcing, such as resonances, are optimally exploited by the resolvent model, whereas they might not be forced as efficiently in the real flow. A method to single out the trapped acoustic wave components in the transonic jet is presented in \citet{SchmidtEtAl_2017_JFM}.

\begin{figure}
	\centerline{\includegraphics[trim=0 0cm 0 5mm, clip, width=1\textwidth]{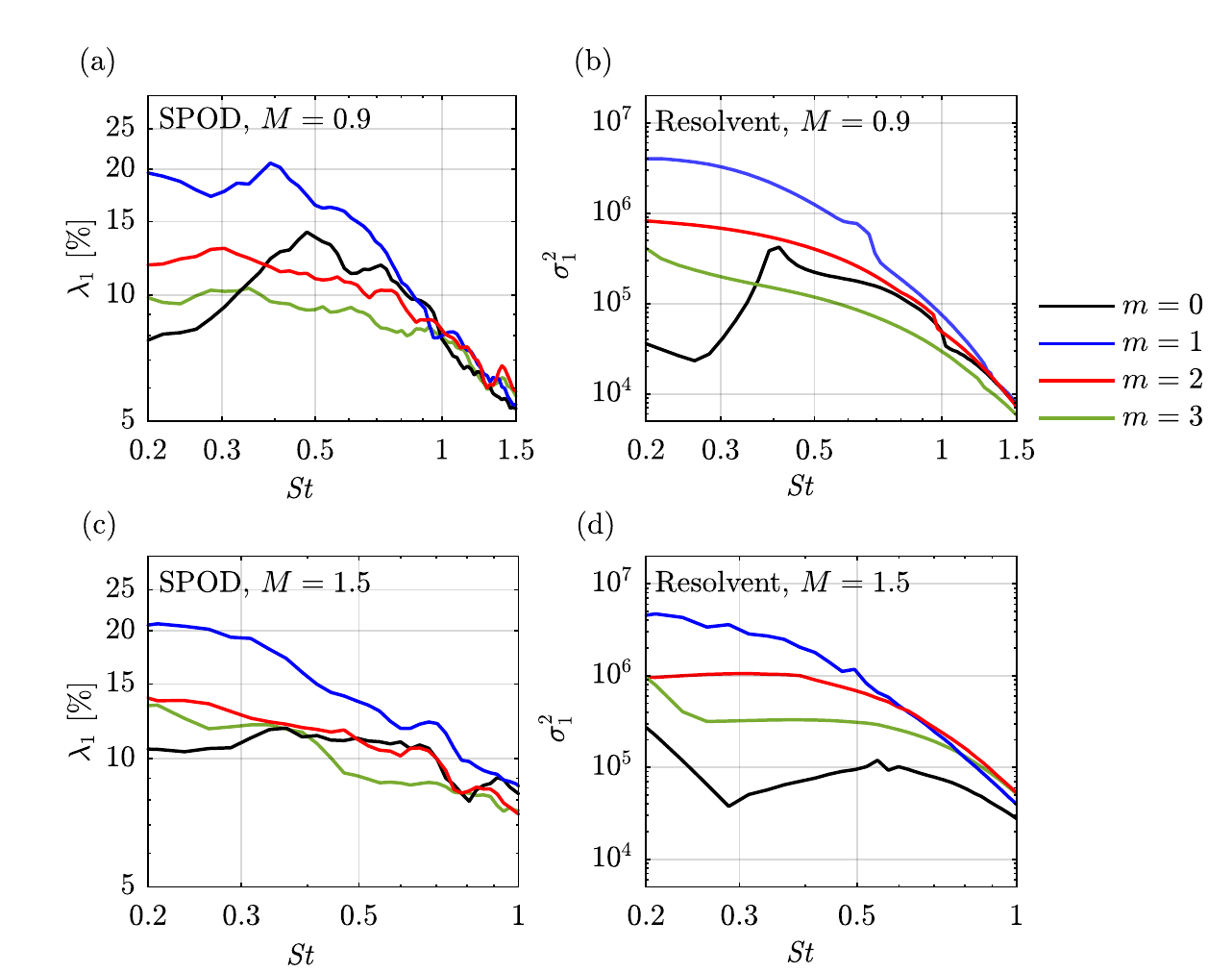}}% Images in 100% size
\captionsetup{singlelinecheck=off}
	\caption[]{SPOD energy spectra (a,c) and optimal resolvent gains (b,d) as in figure \ref{fig:M04_resolventVsSPOD_spectrum_mStudy_globalForcing}, but for the transsonic (a,b) and the supersonic (c,d) jet.}
	\label{fig:M09_resolventVsSPOD_spectrum_mStudy_globalForcing}
\end{figure} 
Figure \ref{fig:M09_resolventVsSPOD_spectrum_mStudy_globalForcing} shows the leading SPOD mode energy and the optimal resolvent gain for the transsonic (top) and supersonic (bottom) cases. As in figure \ref{fig:M04_resolventVsSPOD_spectrum_mStudy_globalForcing}, spectra for the first four azimuthal wavenumbers are reported. Like in the 
subsonic case, favorable agreement of the qualitative trends is found between the SPOD analysis and the resolvent model. The presence of the acoustic resonance mechanism associated with the trapped modes is apparent in the gains. The peaks seen in figure \ref{fig:M04_resolventVsSPOD_spectrum_mStudy_globalForcing}(a), for example at $(\St,m)\approx(0.4,0)$ and $(\St,m)\approx(0.7,1)$, coincide with the frequencies of branches of trapped acoustic modes. The associated eigenvalues are only marginally damped in the global spectra of the same operator \citep{SchmidtEtAl_2017_JFM}. This proximity of the eigenvalues to the real axis explains the peaks in the gain curves as a pseudo-resonance. The acoustic branch locations are marked in the resolvent gain spectra shown in figure \ref{fig:M09_resolventGainVsSt_m01} (appendix \ref{app:mach}).

%%%%%%%%%%%%%%%%%%%%%%%%%%%%%%%%%%%%%%%%%%%%%%%%%%%%%%%%%%%%%%%%%%%%%%%%%%%%%%%%%%%%
\section{Summary and conclusions}\label{discussion}
%%%%%%%%%%%%%%%%%%%%%%%%%%%%%%%%%%%%%%%%%%%%%%%%%%%%%%%%%%%%%%%%%%%%%%%%%%%%%%%%%%%%
Large-scale structures taking the form of spatially modulated wavepackets have long been observed in turbulent jets and past attempts to model them using linear theory have met partial success \citep{jordan2013wave}. In this paper, we use SPOD to distill these wavepackets from a high-fidelity numerical simulation and demonstrate that a resolvent mean flow model predicts them in great detail. Both approaches paint a consistent picture of two coexisting mechanisms. The KH-type instability is active, over a range of frequencies and azimuthal wavenumbers, in the initial shear-layer whereas the region downstream of the close of the potential core is dominated by Orr-type waves. Moreover, in the initial shear layer region, Orr-type waves are also present but not readily observed as they are swamped by the high-gain KH waves. We quantified both types of structures over a range of frequencies and azimuthal wavenumbers. In addition to their differing spatial regions of dominance, they are distinguished by their spatial support, phase speed and frequency scaling. KH-type wavepackets can be regarded as local spatial instabilities. They convect with a phase velocity of $c_{ph}\approx 0.8U_j$ and are triggered by fluctuations close to the nozzle. This spatial separation between optimal forcing and response characterizes a convective non-normality \citep{marquet2009direct} in the presence of a spatial instability mechanism \citep{alizard2009sensitivity,dergham2013stochastic,beneddine2016conditions}. By contrast, Orr-type waves convect at a lower speed in accordance with the jet's velocity-decay rate, and are most effectively sustained by distributed forcing. Both the KH and Orr waves peak at the hight of the critical layer in the radial directions and are optimally forced by the Orr-mechanism.

In the LES data analysis, frequencies at which the KH-mechanism dominates are identified by a separation between the first and second eigenvalues in the SPOD spectrum, and the resolvent gain reliably predicts this low-rank behavior. Although this low-rank behavior can in principal be inferred from the success of past studies based on spatial linear stability theory \citep[e.g.][]{michalke1971instability}, it is most compellingly revealed in the SPOD and resolvent spectra. Equally important to its presence is its absence. For $m=0$ at very low frequencies, for example, the KH-mechanism is absent as the initial shear-layer becomes short as compared to the perturbation wavelength. Here, the jet exhibits non-low-rank behavior and both the SPOD and the resolvent model predict the coexistence of Orr-type waves of similar energy. For $m=1$, on the contrary, the KH-mechanism persists to very low frequencies.

The non-low-rank behavior explains why wavepacket models based on PSE such as the ones by \citet{gudmundsson2011instability}, \citet{cavalieri2013wavepackets} and \citet{sinha2014wavepacket} fail at these the very low frequencies for $m=0$. For example, the PSE method is initialized with the locally most unstable spatial wave, which is then propagated downstream by space-marching. In the non-low-rank regime, this mode does not optimally trigger transient growth and appears as a sub-dominant resolvent mode. Furthermore, as volumetric forcing by the turbulence is not accounted for, PSE cannot support the dominant Orr-type waves. With the goal in mind to further improve its predictive capabilities, in particular at low frequencies, we plan to model the second-order statistics of the forcing and incorporate them into a resolvent-based jet noise model in future work.

\appendix
\section*{Appendix} 
%%%%%%%%%%%%%%%%%%%%%%%%%%%%%%%%%%%%%%%%%%%%%%%%%%%%%%%%%%%%%%%%%%%%%%%%%%%%%%%%%%%%  
\section{Spectral analysis and resolvent model for the transonic and supersonic jets}\label{app:mach}
\begin{figure}
	\centerline{\includegraphics[trim=0 0.1cm 0 6mm, clip, width=1\textwidth]{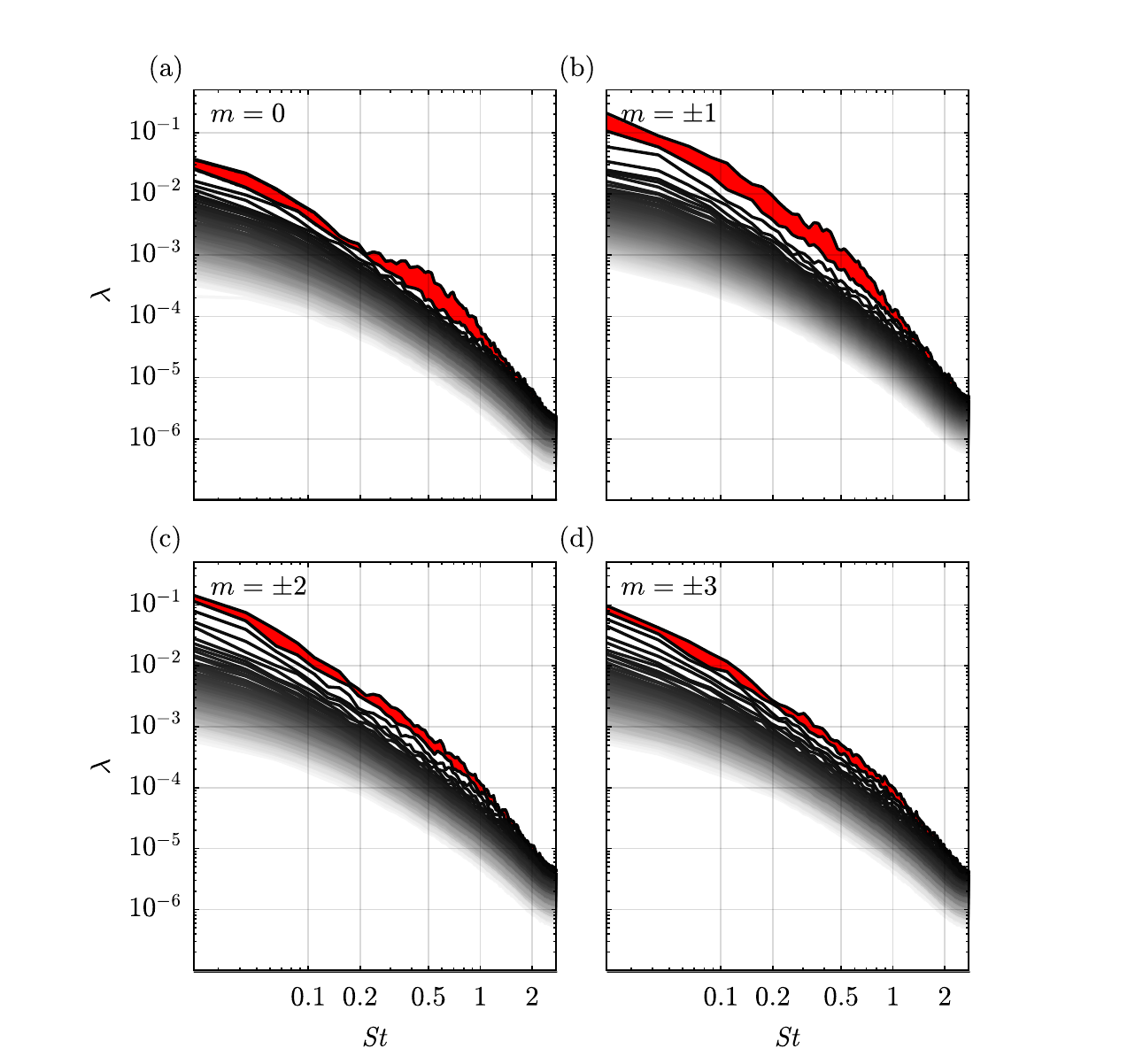}}% Images in 100% size
	\captionsetup{singlelinecheck=off}
	\caption[]{SPOD eigenvalue spectra ($\blacksquare\!{\color{Gray}\blacksquare}\!\square$, $\lambda_1\!>\!\lambda_2\!>\!\cdots\!>\!\lambda_{N}$) as in figure \ref{fig:SPODvsSPODforcing_spectrum_m01}, but for the $M=0.9$ transsonic jet: (a) $m=0$; (b) $m=1$; (c) $m=2$; (d) $m=3$.}
	\label{fig:M09_POD_spectrum_m0123}
\end{figure}

\begin{figure}
	\centerline{\includegraphics[trim=0 0.2cm 0 0mm, clip, width=1\textwidth]{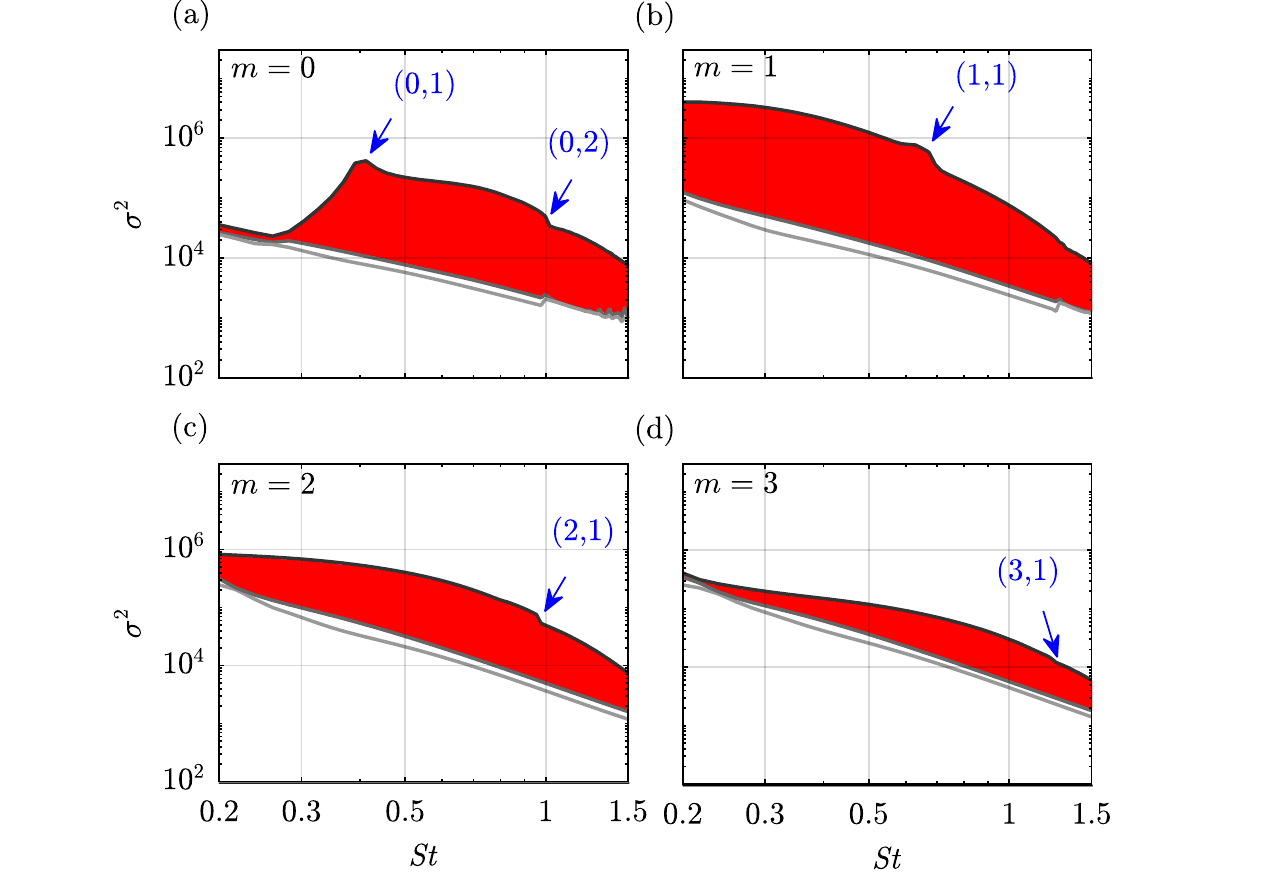}}% Images in 100% size
	\captionsetup{singlelinecheck=off}
	\caption[]{Optimal energetic gain spectra (${\color{Gray}\solidrule}$, $\sigma_{1-3}$) as in figure \ref{fig:M04_gainVsSt_m0123} but for the transonic $M=0.9$ jet: (a) $m=0$; (b) $m=1$; (c) $m=2$; (d) $m=3$. The locations of the branches of trapped acoustic modes are indicated by the doublets $(m,n_r)$, where $n_r$ is their radial order and $m$ the azimuthal wavenumber, as before. See \citep{TowneEtAl_2017_JFM,SchmidtEtAl_2017_JFM} for details.}
	\label{fig:M09_resolventGainVsSt_m01}
\end{figure}

\begin{figure}
	\centerline{\includegraphics[trim=0 0cm 0 6mm, clip, width=1\textwidth]{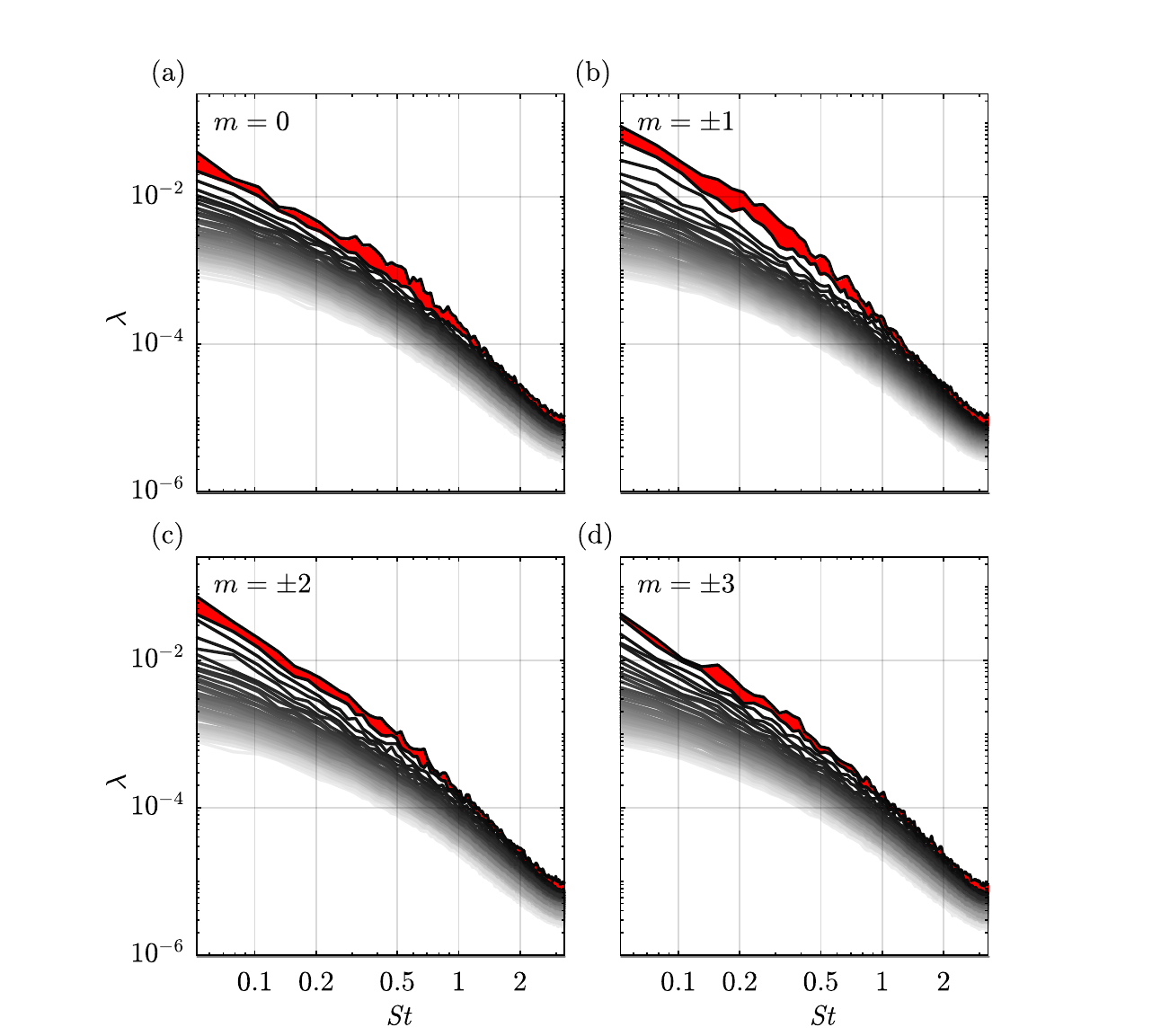}}% Images in 100% size
\captionsetup{singlelinecheck=off}
	\caption[]{SPOD eigenvalue spectra ($\blacksquare\!{\color{Gray}\blacksquare}\!\square$, $\lambda_1\!>\!\lambda_2\!>\!\cdots\!>\!\lambda_{N}$) as in figure \ref{fig:SPODvsSPODforcing_spectrum_m01}, but for the $M=1.5$ supersonic jet: (a) $m=0$; (b) $m=1$; (c) $m=2$; (d) $m=3$.}
	\label{fig:B118_POD_spectrum_m0123}
\end{figure}

\begin{figure}
	\centerline{\includegraphics[width=1\textwidth]{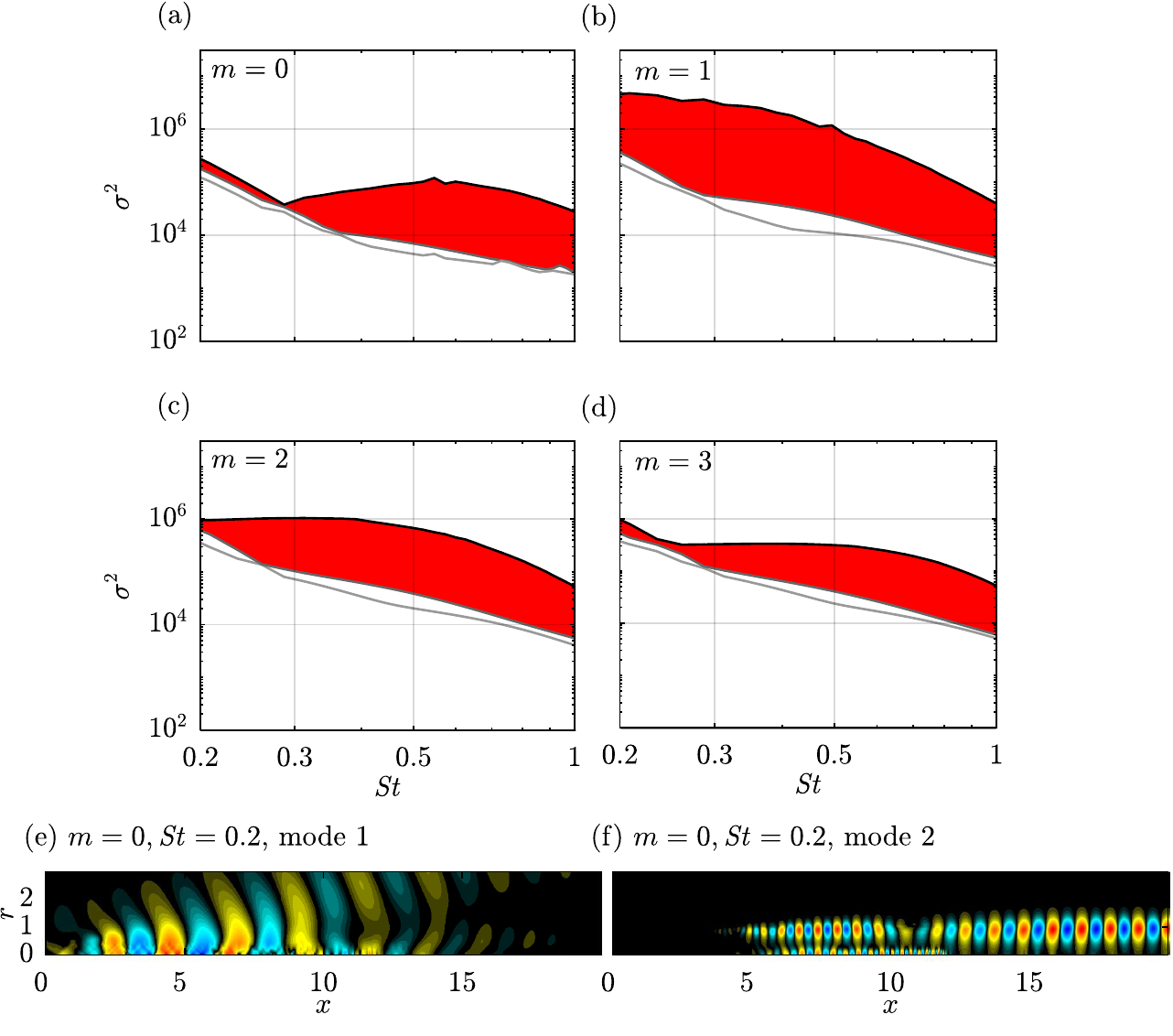}}% Images in 100% size
	\captionsetup{singlelinecheck=off}
	\caption[]{Optimal energetic gain spectra (${\color{Gray}\solidrule}$, $\sigma_{1-3}$) as in figure \ref{fig:M04_gainVsSt_m0123} but for the supersonic $M=1.5$ jet: (a) $m=0$; (b) $m=1$; (c) $m=2$; (d) $m=3$. The pressure of the leading and first suboptimal modes at $(m,\St)=(0,0.2)$ are shown in (e) and (f), respectively.}
	\label{fig:B118_gainVsSt_m0123}
\end{figure}

\begin{figure}
	\centerline{\includegraphics[trim=0 2.7cm 0 1mm, clip, width=1\textwidth]{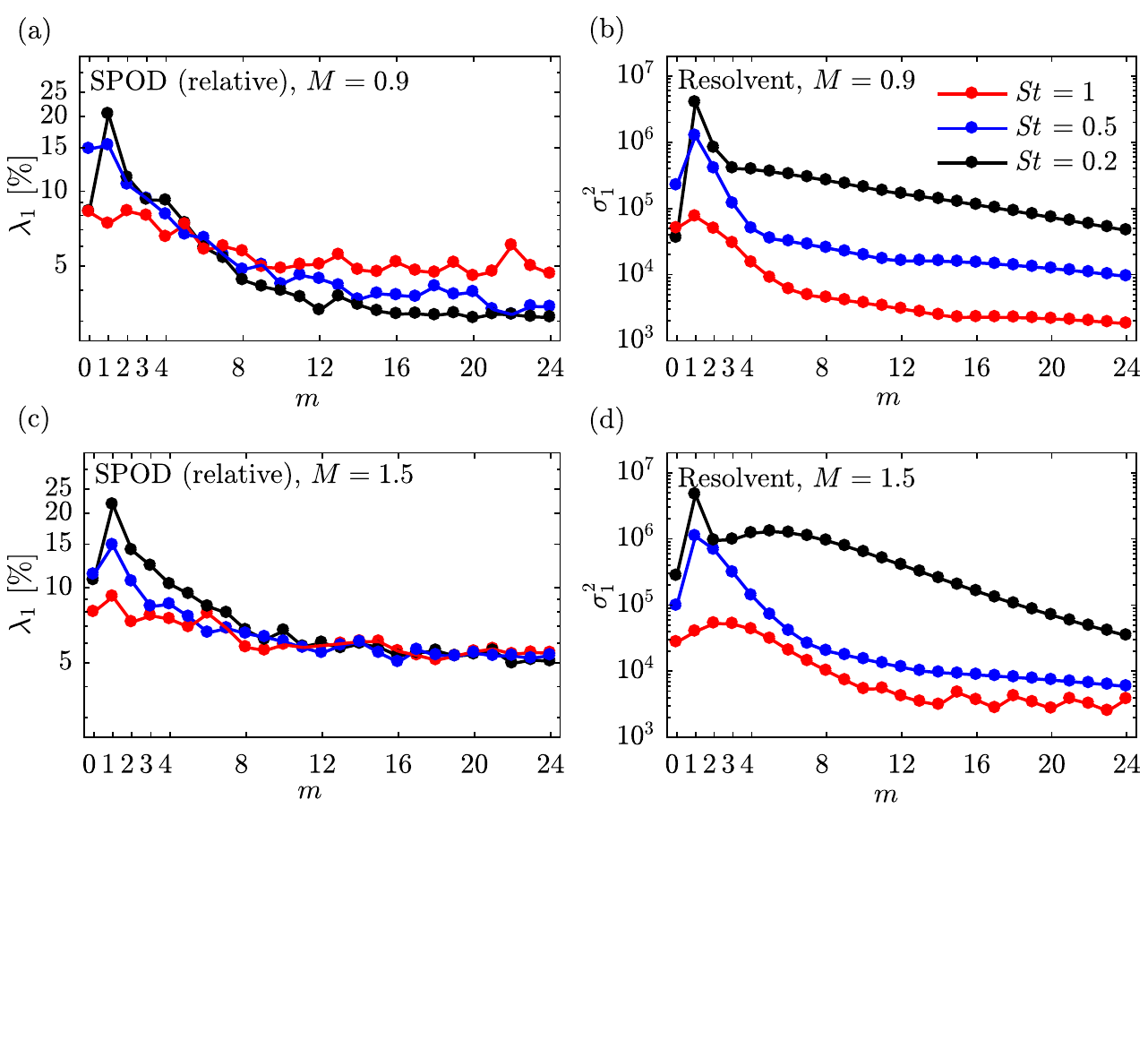}}% Images in 100% size
\captionsetup{singlelinecheck=off}
	\caption[]{Azimuthal wavenumber dependence of the relative modal energy of the SPOD (a,c), and the optimal resolvent gains (b,d) as in figure \ref{fig:M04_resolventAndSPOD_vs_m_percent}, but for the transsonic (a,b), and supersonic jet (c,d).}
	\label{fig:M09andB118_resolventAndSPOD_vs_m_percent}
\end{figure}

%%%%%%%%%%%%%%%%%%%%%%%%%%%%%%%%%%%%%%%%%%%%%%%%%%%%%%%%%%%%%%%%%%%%%%%%%%%%%%%%%%%%   
This appendix reports the additional results for the SPOD and resolvent analyses of the $M=0.9$ and $M=1.5$ jets, that were omitted in \S \ref{mach} for brevity. The resolvent gain spectra for the transonic and supersonic jets are reported in figures \ref{fig:M09_resolventGainVsSt_m01} and \ref{fig:B118_gainVsSt_m0123}, and the SPOD energy spectra in figures \ref{fig:M09_POD_spectrum_m0123} and \ref{fig:B118_POD_spectrum_m0123}, respectively. The azimuthal wavenumber dependence of the optimal gain and SPOD energy is studied in figure \ref{fig:M09andB118_resolventAndSPOD_vs_m_percent} for both jet configurations. 
In figure \ref{fig:M09_resolventGainVsSt_m01}, the locations of the branches of trapped acoustic modes \citep{SchmidtEtAl_2017_JFM} are indicated, and their effect on the resolvent gain becomes apparent.

For the supersonic jet shown in figure \ref{fig:B118_gainVsSt_m0123}, a sudden change of slope is observed in the suboptimal gain curves. An inspection of the modal structures confirms that this change is associated with the presence of upstream-propagating subsonic waves \citep{tam1989three}, as previously discussed in the context of figure \ref{fig:M09vsB118_ResolventVsSPOD_m1}. In panels \ref{fig:B118_gainVsSt_m0123}(e,f), the dominant and the first suboptimal mode for $(m,\St)=(0,0.2)$ are compared. At this frequency, the leading mode is of KH-type, whereas the second mode is of mixed Orr/acoustic type. In the second mode in panel \ref{fig:B118_gainVsSt_m0123}(f), the acoustic wave component appears isolated in the stretch $6\lesssim x\lesssim12$ along the axis. The trapped acoustic waves in the transonic jet and the upstream-propagating subsonic waves in the supersonic jet have a direct effect on the resolvent gain, as can be seen in figures \ref{fig:M09_resolventGainVsSt_m01} and \ref{fig:B118_gainVsSt_m0123}, respectively. It is a remarkable property of the resolvent analysis that it is able to isolate these physical phenomena. Both types of waves are also apparent in the SPOD modes. However, they appear much less pronounces in the latter. This discrepancy is also reflected in the SPOD energy spectra in figures \ref{fig:M09_POD_spectrum_m0123}(a) and \ref{fig:B118_POD_spectrum_m0123}(b), respectively. In the SPOD spectra, the effect of these special waves is not apparent. This observation further highlights the importance of the second-order forcing statistics. Other factors are the Reynolds number dependance and the choice of norm. 

\clearpage

\bibliography{jets,acoustics,others,non-modalStab,mypublications_proceedings,mypublications_journals,books,PSE_and_LST,PODetc}

\begin{thebibliography}{54}
\expandafter\ifx\csname natexlab\endcsname\relax\def\natexlab#1{#1}\fi
\def\au#1{#1} \def\ed#1{#1} \def\yr#1{#1}\def\at#1{#1}\def\jt#1{\textit{#1}}
  \def\bt#1{#1}\def\bvol#1{\textbf{#1}} \def\vol#1{#1} \def\pg#1{#1}
  \def\publ#1{#1}\def\arxiv#1{#1}\def\org#1{#1}\def\st#1{\textit{#1}}

\bibitem[Alizard {\em et~al.\/}(2009)Alizard, Cherubini \&
  Robinet]{alizard2009sensitivity}
{\sc \au{Alizard, F.}, \au{Cherubini, S.} \& \au{Robinet, {J.-C.}}} \yr{2009}
  \at{Sensitivity and optimal forcing response in separated boundary layer
  flows}.  \jt{Physics of Fluids}  \bvol{21}.

\bibitem[Arndt {\em et~al.\/}(1997)Arndt, Long \& Glauser]{arndt1997proper}
{\sc \au{Arndt, R.~E.~A.}, \au{Long, D.~F.} \& \au{Glauser, M.~N.}} \yr{1997}
  \at{The proper orthogonal decomposition of pressure fluctuations surrounding
  a turbulent jet}.  \jt{Journal of Fluid Mechanics}  \bvol{340},  \pg{1--33}.

\bibitem[Aubry(1991)]{aubry1991hidden}
{\sc \au{Aubry, N.}} \yr{1991}  \at{On the hidden beauty of the proper
  orthogonal decomposition}.  \jt{Theoretical and Computational Fluid Dynamics}
   \bvol{2}~(5),  \pg{339--352}.

\bibitem[Beneddine {\em et~al.\/}(2016)Beneddine, Sipp, Arnault, Dandois \&
  Lesshafft]{beneddine2016conditions}
{\sc \au{Beneddine, S.}, \au{Sipp, D.}, \au{Arnault, A.}, \au{Dandois, J.} \&
  \au{Lesshafft, L.}} \yr{2016}  \at{Conditions for validity of mean flow
  stability analysis}.  \jt{Journal of Fluid Mechanics}  \bvol{798},
  \pg{485--504}.

\bibitem[Bishop {\em et~al.\/}(1971)Bishop, Ffowcs~Williams \&
  Smith]{bishop1971noise}
{\sc \au{Bishop, K.~A.}, \au{Ffowcs~Williams, J.~E.} \& \au{Smith, W.}}
  \yr{1971}  \at{On the noise sources of the unsuppressed high-speed jet}.
  \jt{Journal of Fluid Mechanics}  \bvol{50}~(1),  \pg{21--31}.

\bibitem[Br{\`e}s {\em et~al.\/}(2017{\natexlab{{\em a\/}}})Br{\`e}s, Jordan,
  Le~Rallic, Jaunet, Cavalieri, Towne, Lele, Colonius \& Schmidt]{bres2017jets}
{\sc \au{Br{\`e}s, G.}, \au{Jordan, P.}, \au{Le~Rallic, M.}, \au{Jaunet, V.},
  \au{Cavalieri, A.~V.~G}, \au{Towne, A.}, \au{Lele, S.}, \au{Colonius, T.} \&
  \au{Schmidt, O.~T.}} \yr{2017{\natexlab{{\em a\/}}}}  \at{Importance of the
  nozzle-exit boundary-layer state in subsonic turbulent jets}.  \jt{submitted
  to Journal of Fluid Mechanics} .

\bibitem[Br{\`e}s {\em et~al.\/}(2017{\natexlab{{\em b\/}}})Br{\`e}s, Ham,
  Nichols \& Lele]{bres2017unstructured}
{\sc \au{Br{\`e}s, G.~A.}, \au{Ham, F.~E.}, \au{Nichols, J.~W.} \& \au{Lele,
  S.~K.}} \yr{2017{\natexlab{{\em b\/}}}}  \at{Unstructured large-eddy
  simulations of supersonic jets}.  \jt{AIAA Journal}  \bvol{55}~(4),
  \pg{1164--1184}.

\bibitem[Cavalieri {\em et~al.\/}(2011)Cavalieri, Jordan, Agarwal \&
  Gervais]{cavalieri2011jittering}
{\sc \au{Cavalieri, A.~V.~G.}, \au{Jordan, P.}, \au{Agarwal, A.} \&
  \au{Gervais, Y.}} \yr{2011}  \at{Jittering wave-packet models for subsonic
  jet noise}.  \jt{Journal of Sound and Vibration}  \bvol{330}~(18),
  \pg{4474--4492}.

\bibitem[Cavalieri {\em et~al.\/}(2013)Cavalieri, Rodr\'{i}guez, Jordan,
  Colonius \& Gervais]{cavalieri2013wavepackets}
{\sc \au{Cavalieri, A.~V.~G}, \au{Rodr\'{i}guez, D.}, \au{Jordan, P.},
  \au{Colonius, T.} \& \au{Gervais, Y.}} \yr{2013}  \at{Wavepackets in the
  velocity field of turbulent jets}.  \jt{Journal of Fluid Mechanics}
  \bvol{730},  \pg{559--592}.

\bibitem[Cavalieri {\em et~al.\/}(2016)Cavalieri, Sasaki, Jordan, Schmidt,
  Colonius \& Brs]{CavalieriSasakiJordanEtAl2016}
{\sc \au{Cavalieri, A.~V.~G.}, \au{Sasaki, K.}, \au{Jordan, P.}, \au{Schmidt,
  O.~T.}, \au{Colonius, T.} \& \au{Brs, G.~A.}} \yr{2016} High-frequency
  wavepackets in turbulent jets.  \bt{In {\em 22nd AIAA/CEAS Aeroacoustics
  Conference\/}}.  \publ{American Institute of Aeronautics and Astronautics
  (AIAA)}.

\bibitem[Chu(1965)]{chu1965energy}
{\sc \au{Chu, {B.-T.}}} \yr{1965}  \at{On the energy transfer to small
  disturbances in fluid flow ({P}art {I})}.  \jt{Acta Mechanica}  \bvol{1}~(3),
   \pg{215--234}.

\bibitem[Citriniti \& George(2000)]{citriniti2000reconstruction}
{\sc \au{Citriniti, J.~H.} \& \au{George, W.~K.}} \yr{2000}  \at{Reconstruction
  of the global velocity field in the axisymmetric mixing layer utilizing the
  proper orthogonal decomposition}.  \jt{Journal of Fluid Mechanics}
  \bvol{418},  \pg{137--166}.

\bibitem[Crighton \& Gaster(1976)]{crighton1976stability}
{\sc \au{Crighton, D.~G.} \& \au{Gaster, M.}} \yr{1976}  \at{Stability of
  slowly diverging jet flow}.  \jt{Journal of Fluid Mechanics}  \bvol{77}~(02),
   \pg{397--413}.

\bibitem[Crighton \& Huerre(1990)]{crighton1990shear}
{\sc \au{Crighton, D.~G.} \& \au{Huerre, P.}} \yr{1990}  \at{Shear-layer
  pressure fluctuations and superdirective acoustic sources}.  \jt{Journal of
  Fluid Mechanics}  \bvol{220},  \pg{355--368}.

\bibitem[Crow \& Champagne(1971)]{crow1971orderly}
{\sc \au{Crow, S.~C.} \& \au{Champagne, F.~H.}} \yr{1971}  \at{Orderly
  structure in jet turbulence}.  \jt{Journal of Fluid Mechanics}
  \bvol{48}~(03),  \pg{547--591}.

\bibitem[Dergham {\em et~al.\/}(2013)Dergham, Sipp \&
  Robinet]{dergham2013stochastic}
{\sc \au{Dergham, G.}, \au{Sipp, D.} \& \au{Robinet, J-C.}} \yr{2013}
  \at{Stochastic dynamics and model reduction of amplifier flows: the backward
  facing step flow}.  \jt{Journal of Fluid Mechanics}  \bvol{719},
  \pg{406--430}.

\bibitem[Farrell \& Ioannou(1993)]{farrell1993stochastic}
{\sc \au{Farrell, B.~F.} \& \au{Ioannou, P.~J.}} \yr{1993}  \at{Stochastic
  forcing of the linearized navier--stokes equations}.  \jt{Physics of Fluids
  A: Fluid Dynamics (1989-1993)}  \bvol{5}~(11),  \pg{2600--2609}.

\bibitem[Garnaud {\em et~al.\/}(2013)Garnaud, Lesshafft, Schmid \&
  Huerre]{garnaud2013pref}
{\sc \au{Garnaud, X.}, \au{Lesshafft, L.}, \au{Schmid, P.~J.} \& \au{Huerre,
  P.}} \yr{2013}  \at{The preferred mode of incompressible jets: linear
  frequency response analysis}.  \jt{Journal of Fluid Mechanics}  \bvol{716},
  \pg{189--202}.

\bibitem[Glauser {\em et~al.\/}(1987)Glauser, Leib \&
  George]{glauser1987coherent}
{\sc \au{Glauser, M.~N.}, \au{Leib, S.~J.} \& \au{George, W.~K.}} \yr{1987}
  \at{Coherent structures in the axisymmetric turbulent jet mixing layer}.
  \jt{Turbulent shear flows}  \bvol{5},  \pg{134--145}.

\bibitem[Gudmundsson \& Colonius(2011)]{gudmundsson2011instability}
{\sc \au{Gudmundsson, K.} \& \au{Colonius, T.}} \yr{2011}  \at{Instability wave
  models for the near-field fluctuations of turbulent jets}.  \jt{Journal of
  Fluid Mechanics}  \bvol{689},  \pg{97--128}.

\bibitem[Jeun {\em et~al.\/}(2016)Jeun, Nichols \&
  Jovanovi{\'c}]{jeun2016input}
{\sc \au{Jeun, J.}, \au{Nichols, J.~W.} \& \au{Jovanovi{\'c}, M.~R.}} \yr{2016}
   \at{Input-output analysis of high-speed axisymmetric isothermal jet noise}.
  \jt{Physics of Fluids (1994-present)}  \bvol{28}~(4),  \pg{047101}.

\bibitem[Jordan \& Colonius(2013)]{jordan2013wave}
{\sc \au{Jordan, P.} \& \au{Colonius, T.}} \yr{2013}  \at{Wave packets and
  turbulent jet noise}.  \jt{Annual Review of Fluid Mechanics}  \bvol{45},
  \pg{173--195}.

\bibitem[Jordan {\em et~al.\/}(2017)Jordan, Zhang, Lehnasch \&
  Cavalieri]{jordan2017modal}
{\sc \au{Jordan, P.}, \au{Zhang, M.}, \au{Lehnasch, G.} \& \au{Cavalieri,
  A.~V.~G.}} \yr{2017} Modal and non-modal linear wavepacket dynamics in
  turbulent jets.  \bt{In {\em 23rd AIAA/CEAS Aeroacoustics Conference\/}},
  \pg{p. 3379}.

\bibitem[Jovanovi\'c \& Bamieh(2005)]{jovanovic2005componentwise}
{\sc \au{Jovanovi\'c, M.~R.} \& \au{Bamieh, B.}} \yr{2005}  \at{Componentwise
  energy amplification in channel flows}.  \jt{Journal of Fluid Mechanics}
  \bvol{534},  \pg{145--183}.

\bibitem[Lumley(1967)]{Lumley:1967}
{\sc \au{Lumley, J.~L.}} \yr{1967}  \at{The structure of inhomogeneous
  turbulent flows}.  \bt{In {\em Atmospheric turbulence and radio
  propagation\/} (ed. \ed{A.~M. Yaglom \& V.~I. Tatarski})},  \pg{pp.
  166--178}.  \publ{Moscow: Nauka}.

\bibitem[Lumley(1970)]{Lumley:1970}
{\sc \au{Lumley, J.~L.}} \yr{1970} {\em Stochastic tools in turbulence\/}.
  \publ{New York: Academic Press}.

\bibitem[Marquet {\em et~al.\/}(2009)Marquet, Lombardi, Chomaz, Sipp \&
  Jacquin]{marquet2009direct}
{\sc \au{Marquet, O.}, \au{Lombardi, M.}, \au{Chomaz, J.-M.}, \au{Sipp, D.} \&
  \au{Jacquin, L.}} \yr{2009}  \at{Direct and adjoint global modes of a
  recirculation bubble: lift-up and convective non-normalities}.  \jt{Journal
  of Fluid Mechanics}  \bvol{622},  \pg{1--21}.

\bibitem[McKeon \& Sharma(2010)]{McKeonSharma2010}
{\sc \au{McKeon, B.~J.} \& \au{Sharma, A.~S.}} \yr{2010}  \at{A critical-layer
  framework for turbulent pipe flow}.  \jt{Journal of Fluid Mechanics}
  \bvol{658},  \pg{336–382}.

\bibitem[Mettot {\em et~al.\/}(2014)Mettot, Sipp \&
  B{\'e}zard]{mettot2014quasi}
{\sc \au{Mettot, C.}, \au{Sipp, D.} \& \au{B{\'e}zard, H.}} \yr{2014}
  \at{Quasi-laminar stability and sensitivity analyses for turbulent flows:
  Prediction of low-frequency unsteadiness and passive control}.  \jt{Physics
  of Fluids (1994-present)}  \bvol{26}~(4),  \pg{045112}.

\bibitem[Michalke(1971)]{michalke1971instability}
{\sc \au{Michalke, A.}} \yr{1971}  \at{Instability of a compressible circular
  free jet with consideration of the influence of the jet boundary layer
  thickness}.  \jt{Z. f{\"u}r Flugwissenschaften}  \bvol{19}~(8),
  \pg{319--328}.

\bibitem[Moarref \& Jovanovi{\'c}(2012)]{moarref2012model}
{\sc \au{Moarref, R.} \& \au{Jovanovi{\'c}, M.~R.}} \yr{2012}  \at{Model-based
  design of transverse wall oscillations for turbulent drag reduction}.
  \jt{Journal of Fluid Mechanics}  \bvol{707},  \pg{205--240}.

\bibitem[Monokrousos {\em et~al.\/}(2010)Monokrousos, \AA{}kervik, Brandt \&
  Henningson]{monokrousos2010global}
{\sc \au{Monokrousos, A.}, \au{\AA{}kervik, E.}, \au{Brandt, L.} \&
  \au{Henningson, D.~S.}} \yr{2010}  \at{Global three-dimensional optimal
  disturbances in the blasius boundary-layer flow using time-steppers}.
  \jt{Journal of Fluid Mechanics}  \bvol{650},  \pg{181--214}.

\bibitem[Moore(1977)]{moore1977role}
{\sc \au{Moore, C.~J.}} \yr{1977}  \at{The role of shear-layer instability
  waves in jet exhaust noise}.  \jt{Journal of Fluid Mechanics}
  \bvol{80}~(02),  \pg{321--367}.

\bibitem[Pope(2000)]{Pope2000}
{\sc \au{Pope, S.~B.}} \yr{2000} {\em Turbulent Flows\/}, 1st edn.
  \publ{Cambridge University Press}.

\bibitem[Reddy \& Henningson(1993)]{reddy1993energy}
{\sc \au{Reddy, S.~C.} \& \au{Henningson, D.~S.}} \yr{1993}  \at{Energy growth
  in viscous channel flows}.  \jt{Journal of Fluid Mechanics}  \bvol{252}~(1),
  \pg{209--238}.

\bibitem[Reddy {\em et~al.\/}(1993)Reddy, Schmid \&
  Henningson]{reddy1993pseudospectra}
{\sc \au{Reddy, S.~C.}, \au{Schmid, P.~J.} \& \au{Henningson, D.~S.}} \yr{1993}
   \at{Pseudospectra of the {O}rr-{S}ommerfeld operator}.  \jt{SIAM Journal on
  Applied Mathematics}  \bvol{53}~(1),  \pg{15--47}.

\bibitem[Sasaki {\em et~al.\/}(2017)Sasaki, Cavalieri, Jordan, Schmidt,
  Colonius \& Brès]{SasakiEtAl2017}
{\sc \au{Sasaki, K.}, \au{Cavalieri, A.~V.~G.}, \au{Jordan, P.}, \au{Schmidt,
  O.~T.}, \au{Colonius, T.} \& \au{Brès, G.~A.}} \yr{2017}  \at{High-frequency
  wavepackets in turbulent jets}.  \jt{Journal of Fluid Mechanics}  \bvol{830},
   \pg{R2}.

\bibitem[Schlinker {\em et~al.\/}(2009)Schlinker, Simonich, Shannon, Reba,
  Colonius, Gudmundsson \& Ladeinde]{SchlinkerEtAl_2009_AIAA}
{\sc \au{Schlinker, R.~H.}, \au{Simonich, J.~C.}, \au{Shannon, D.~W.},
  \au{Reba, R.~A.}, \au{Colonius, T.}, \au{Gudmundsson, K.} \& \au{Ladeinde,
  F.}} \yr{2009}  \at{Supersonic jet noise from round and chevron nozzles:
  experimental studies}.  \jt{AIAA paper}  \bvol{3257},  \pg{2009}.

\bibitem[Schmid \& Henningson(2001)]{schmid2001stability}
{\sc \au{Schmid, P.J.} \& \au{Henningson, D.~S.}} \yr{2001} {\em Stability and
  Transition in Shear Flows\/}, 1st edn.  \publ{Springer-Verlag New York}.

\bibitem[{Schmid}(2010)]{schmid2010dmd}
{\sc \au{{Schmid}, P.~J.}} \yr{2010}  \at{{Dynamic mode decomposition of
  numerical and experimental data}}.  \jt{Journal of Fluid Mechanics}
  \bvol{656},  \pg{5--28}.

\bibitem[Schmidt {\em et~al.\/}(2017)Schmidt, Towne, Colonius, Cavalieri,
  Jordan \& Brès]{SchmidtEtAl_2017_JFM}
{\sc \au{Schmidt, O.~T.}, \au{Towne, A.}, \au{Colonius, T.}, \au{Cavalieri,
  A.~V.~G.}, \au{Jordan, P.} \& \au{Brès, G.~A.}} \yr{2017}  \at{Wavepackets
  and trapped acoustic modes in a turbulent jet: coherent structure eduction
  and global stability}.  \jt{Journal of Fluid Mechanics}  \bvol{825},
  \pg{1153–1181}.

\bibitem[Semeraro {\em et~al.\/}(2016)Semeraro, Lesshafft, Jaunet \&
  Jordan]{lesshafftmodeling}
{\sc \au{Semeraro, O.}, \au{Lesshafft, L.}, \au{Jaunet, V.} \& \au{Jordan, P.}}
  \yr{2016}  \at{Modeling of coherent structures in a turbulent jet as global
  linear instability wavepackets: Theory and experiment}.  \jt{International
  Journal of Heat and Fluid Flow}  \bvol{62},  \pg{24--32}.

\bibitem[Sinha {\em et~al.\/}(2014)Sinha, Rodr{\'\i}guez, Br{\`e}s \&
  Colonius]{sinha2014wavepacket}
{\sc \au{Sinha, A.}, \au{Rodr{\'\i}guez, D.}, \au{Br{\`e}s, G.~A.} \&
  \au{Colonius, T.}} \yr{2014}  \at{Wavepacket models for supersonic jet
  noise}.  \jt{Journal of Fluid Mechanics}  \bvol{742},  \pg{71--95}.

\bibitem[Sipp \& Marquet(2013)]{sipp2012characterization}
{\sc \au{Sipp, D.} \& \au{Marquet, O.}} \yr{2013}  \at{Characterization of
  noise amplifiers with global singular modes: the case of the leading-edge
  flat-plate boundary layer}.  \jt{Theoretical and Computational Fluid
  Dynamics}  \bvol{27}~(5),  \pg{617--635}.

\bibitem[Sirovich(1987)]{sirovich1987turbulence}
{\sc \au{Sirovich, L.}} \yr{1987}  \at{Turbulence and the dynamics of coherent
  structures}.  \jt{Quarterly of applied mathematics}  \bvol{45}~(3),
  \pg{561--571}.

\bibitem[Suzuki \& Colonius(2006)]{suzuki2006instability}
{\sc \au{Suzuki, T.} \& \au{Colonius, T.}} \yr{2006}  \at{Instability waves in
  a subsonic round jet detected using a near-field phased microphone array}.
  \jt{Journal of Fluid Mechanics}  \bvol{565}~(1),  \pg{197--226}.

\bibitem[Tam \& Hu(1989)]{tam1989three}
{\sc \au{Tam, C.~K.~W.} \& \au{Hu, F.~Q.}} \yr{1989}  \at{On the three families
  of instability waves of high-speed jets}.  \jt{Journal of Fluid Mechanics}
  \bvol{201},  \pg{447--483}.

\bibitem[Tissot {\em et~al.\/}(2017)Tissot, Zhang, Laj{\' u}s, Cavalieri \&
  Jordan]{tissot2017}
{\sc \au{Tissot, G.}, \au{Zhang, M.}, \au{Laj{\' u}s, F.~C.}, \au{Cavalieri,
  A.~V.~G.} \& \au{Jordan, P.}} \yr{2017}  \at{Sensitivity of wavepackets in
  jets to nonlinear effects: the role of the critical layer}.  \jt{Journal of
  Fluid Mechanics}  \bvol{811},  \pg{95--137}.

\bibitem[Towne {\em et~al.\/}(2017{\natexlab{{\em a\/}}})Towne, Br\`es \&
  Lele]{TowneEtAl_2017_AIAA}
{\sc \au{Towne, A.}, \au{Br\`es, G.~A.} \& \au{Lele, S.~K.}}
  \yr{2017{\natexlab{{\em a\/}}}}  \at{A statistical jet-noise model based on
  the resolvent framework}.  \jt{AIAA Paper $\#$2017-3406} .

\bibitem[Towne {\em et~al.\/}(2017{\natexlab{{\em b\/}}})Towne, Cavalieri,
  Jordan, Colonius, Schmidt, Jaunet \& Brès]{TowneEtAl_2017_JFM}
{\sc \au{Towne, A.}, \au{Cavalieri, A.~V.~G.}, \au{Jordan, P.}, \au{Colonius,
  T.}, \au{Schmidt, O.~T.}, \au{Jaunet, V.} \& \au{Brès, G.~A.}}
  \yr{2017{\natexlab{{\em b\/}}}}  \at{Acoustic resonance in the potential core
  of subsonic jets}.  \jt{Journal of Fluid Mechanics}  \bvol{825},
  \pg{1113–1152}.

\bibitem[Towne {\em et~al.\/}(2015)Towne, Colonius, Jordan \&
  Br{\`e}s]{towne2015stochastic}
{\sc \au{Towne, A.}, \au{Colonius, T.}, \au{Jordan, P., Cavalieri~A.~V.~G.} \&
  \au{Br{\`e}s, G.~A.}} \yr{2015} Stochastic and nonlinear forcing of
  wavepackets in a mach 0.9 jet.  \bt{In {\em 21st AIAA/CEAS Aeroacoustics
  Conference\/}},  \pg{p. 2217}.

\bibitem[Towne {\em et~al.\/}(2017{\natexlab{{\em c\/}}})Towne, Schmidt \&
  Colonius]{TowneEtAt2017SPOD}
{\sc \au{Towne, A.}, \au{Schmidt, O.~T.} \& \au{Colonius, T.}}
  \yr{2017{\natexlab{{\em c\/}}}}  \at{Spectral proper orthogonal decomposition
  and its relationship to dynamic mode decomposition and resolvent analysis}.
  \jt{arXiv preprint arXiv:1708.04393} .

\bibitem[Trefethen {\em et~al.\/}(1993)Trefethen, Trefethen, Reddy \&
  Driscoll]{trefethen1993hydrodynamic}
{\sc \au{Trefethen, L.~N.}, \au{Trefethen, A.~E.}, \au{Reddy, S.~C.} \&
  \au{Driscoll, T.~A.}} \yr{1993}  \at{Hydrodynamic stability without
  eigenvalues}.  \jt{Science}  \bvol{261}~(5121),  \pg{578--584}.

\bibitem[Zare {\em et~al.\/}(2017)Zare, Jovanovi{\'c} \&
  Georgiou]{ZareEtAl_2017_JFM}
{\sc \au{Zare, A.}, \au{Jovanovi{\'c}, M.~R.} \& \au{Georgiou, T.~T.}}
  \yr{2017}  \at{Colour of turbulence}.  \jt{Journal of Fluid Mechanics}
  \bvol{812},  \pg{636--680}.

\end{thebibliography}
\bibliographystyle{jfm}

\end{document}